\documentclass[twocolumn]{aastex62}
\usepackage{float,graphicx,amsmath,multirow,mathtools}
\usepackage{color}
\usepackage[version=4]{mhchem}
\usepackage{comment}
\usepackage{amsmath}
\usepackage[caption=false]{subfig}
\usepackage{comment}

\usepackage{bm}

\newcommand{\kms}{km~s$^{-1}$}

\newcommand{\cmnegtwo}{cm$^{-2}$}
\newcommand{\cmnegthree}{cm$^{-3}$}
\newcommand{\hms}[3]{#1\mathrm{h}#2\mathrm{m}#3\mathrm{s}}
\newcommand{\dms}[3]{#1\degr#2\arcmin#3\arcsec}
\begin{document}

\title{A Search for Light Hydrides in the Envelopes of Evolved Stars}
\author{Mark A. Siebert}
\affiliation{Department of Astronomy, University of Virginia, Charlottesville, VA 22904, USA}
\author{Ignacio Simon}
\affiliation{National Radio Astronomy Observatory, Charlottesville, VA 22903, USA}
\author{Christopher N. Shingledecker}
\affiliation{Department of Physics and Astronomy, Benedictine College, Atchison, KS 66002, USA}
\affiliation{Center for Astrochemical Studies, Max-Planck-Institute for Extraterrestrial Physics, Giessenbachstrasse 1, 85748 Garching, Germany}
\affiliation{Institute for Theoretical Chemistry, University Stuttgart, Pfaffenwaldring 55, 70569 Stuttgart, Germany}
\author{P. Brandon Carroll}
\affiliation{Harvard-Smithsonian Center for Astrophysics, Cambridge, MA 02138, USA}
\author{Andrew M. Burkhardt}
\affiliation{Harvard-Smithsonian Center for Astrophysics, Cambridge, MA 02138, USA}
\affiliation{Department of Astronomy, University of Virginia, Charlottesville, VA 22904, USA}
\author{Shawn Thomas Booth}
\affiliation{National Radio Astronomy Observatory, Charlottesville, VA 22903, USA}
\author{Anthony J. Remijan}
\affiliation{National Radio Astronomy Observatory, Charlottesville, VA 22903, USA}

\author{Rebeca Aladro}
\affiliation{Max-Planck-Institut f\"ur Radioastronomie, Auf dem H\"ugel 69, 53121 Bonn, Germany}
\author{Carlos A. Duran}
\affiliation{Max-Planck-Institut f\"ur Radioastronomie, Auf dem H\"ugel 69, 53121 Bonn, Germany}
\author{Brett A. McGuire}
\affiliation{Department of Chemistry, Massachusetts Institute of Technology, Cambridge, MA 02139, USA}
\affiliation{National Radio Astronomy Observatory, Charlottesville, VA 22903, USA}
\affiliation{Harvard-Smithsonian Center for Astrophysics, Cambridge, MA 02138, USA}
\correspondingauthor{Mark A. Siebert, Brett A. McGuire}
\email{mas5fb@virginia.edu, brettmc@mit.edu}

\accepted{August 2, 2020}
\submitjournal{ApJ}

\begin{abstract}

We report a search for the diatomic hydrides SiH, PH, and FeH along the line of sight toward the chemically rich circumstellar envelopes of IRC+10216 and VY Canis Majoris. These molecules are thought to form in high temperature regions near the photospheres of these stars, and may then further react via gas-phase and dust-grain interactions leading to more complex species, but have yet to be constrained by observation. We used the GREAT spectrometer on SOFIA to search for rotational emission lines of these molecules in four spectral windows ranging from 600 GHz to 1500 GHz. Though none of the targeted species were detected in our search, we report their upper limit abundances in each source and discuss how they influence the current understanding of hydride chemistry in dense circumstellar media. We attribute the non-detections of these hydrides to their compact source sizes, high barriers of formation, and proclivity to react with other molecules in the winds.

\end{abstract}

\section{Introduction}
\label{intro}

The circumstellar material of a star in its final stages of evolution is well-known for its unique and complex chemical inventory. These objects form around Asymptotic Giant Branch (AGB) and Red Supergiant (RSG) stars when interior thermal pulses eject loosely bound surface material in the form of stellar winds \citep{Ziurys:2006n}. These steady winds form large-scale spherical structures of gas and dust known as circumstellar envelopes (CSEs). Due to their large range of temperatures and densities, as well as an increased abundance of elements produced through nucleosynthesis, CSEs exhibit incredibly diverse chemistry. Of the 204 unique molecules detected in space, more than 60 were first discovered in these objects \citep{2018ApJS..239...17M}.

Because the physical conditions near the photospheres of these stars are favorable for the formation of silicate and carbonaceous condensates, AGB and RSG stars are well-known for their efficient production of dust grains. \cite{1989IAUS..135..445G} showed that more than $90\%$ of interstellar dust in the galaxy is produced in the envelopes of evolved stars. After they form, dust grains can be studied in great detail through infrared and X-ray spectroscopy, as well as direct analysis of interplanetary material. Through these methods, it has been shown that grains are composed of both crystalline and amorphous magnesium-rich silicates (e.g. olivine, pyroxene, etc.) as well as carbonaceous structures like graphite and amorphous silicon carbide (SiC) and titanium carbide (TiC) clusters. \citep{1989ApJ...341.1059C, 2019A&A...630A.143R,1998A&A...331L..61W,1990Natur.345..238A,1991ApJ...373L..73B}.

Though the evolution and interactions of micron-sized dust grains are largely understood in the ISM, the gas-phase chemical pathways leading to their formation are not well-constrained by the present body of observational work. Thus, characterizing the reservoir of molecules in the dust production zones of evolved stellar envelopes (5--20 stellar radii) is crucial to understanding the structure of interstellar grains \citep{Ziurys:2006n}.

One class of molecules that is expected to form in these hot, dense environments are diatomic hydrides (XH). These species have been intensely studied since the discovery of CH - the first detected interstellar molecule - in 1937 \citep{1937ApJ....86..483S}. Since then, discoveries of OH, HCl, NH, HF, SH, and (tentatively) SiH in the interstellar medium have added to our understanding of hydrides in space \citep{Adams:1941tj,1985ApJ...295..501B,1991ApJ...376L..49M,1997ApJ...488L.141N,2012A&A...542L...6N}. Thermochemical equilibrium models predict a wide variety of these hydride species abundant in the inner winds of evolved stars \citep{2020A&A...637A..59A}. However, only three (OH, HCl, and HF) have been detected in these environments to date \citep{2011A&A...533L...6A,Cernicharo:2010cv}. Thus, searching for other monohydrides in CSEs presents a useful opportunity to better understand the inner regions of these important astronomical sources.

Previous theoretical and observational studies of these objects suggest that the ideal candidates for such a search include silicon monohydride (\ce{SiH}), phosphinidene (\ce{PH}), and iron hydride (\ce{FeH}). All these species are predicted to form in the hot inner regions of CSEs through gas-phase reactions of atomic Si, P, and Fe with neutral hydrogen \citep{2020A&A...637A..59A,Cherchneff:2012ep}, and therefore could exist further out in the envelope as well. Furthermore, \citet{carroll_spectrum_1972} simultaneously reported both the laboratory and solar identification of FeH through observation of its spectrum in the region between 2360~--~8900~\AA. This detection was later confirmed by \citet{carroll_iron_1976}. Around the same time, the Wing-Ford band at 9900~\AA$\;$ was also identified as arising from FeH \citep{wing_infrared_1969,nordh_proposed_1977,wing_confirmation_1977,mould_iron_1978}. It is now known that FeH can form in the atmospheres of stars of spectral types M, S, and K \citep{Mould:1978}, but despite this, it has not yet been detected in the molecular winds of evolved stars.

Though neither \ce{SiH} nor \ce{PH} has been detected in such a CSE to date, their heavier counterparts, silane (\ce{SiH4}) and phosphine (\ce{PH3}), have both been observed in CSEs, and most proposed formation mechanisms for these molecules involve \ce{SiH} and \ce{PH} as an intermediate step \citep{2000ApJ...543..868M,2014ApJ...790L..27A}. Therefore, constraining the abundances of the molecules targeted in this work could not only provide a more complete understanding of the dust formation zones in these objects, but it could also shed light on the chemical processes that lead to the formation of more complex hydrides in CSEs.

Because these are very light molecules, most of their rotational transitions fall in the THz to far-infrared regions of the electromagnetic spectrum. This makes them ideal targets for the German Receiver for Astronomy at Terahertz Frequencies (GREAT) aboard the Stratospheric Observatory for Infrared Astronomy (SOFIA), which operates from 0.5--4.8 THz at an altitude of 12 km, largely eliminating atmospheric absorption due to water vapor. We used this instrument to conduct a search for SiH, PH, and FeH toward two well-studied CSEs. 

We observed the carbon star IRC+10216 to search for these hydrides in a circumstellar environment dominated by carbon chemistry. This source has been the subject of numerous single-dish and interferometric spectral line surveys \citep{Avery:1992A,Zhang:2017A,Cernicharo:2013A}, revealing a rich gas-phase chemical inventory. Currently, the only diatomic hydrides observed toward IRC+10216 are the halide bearing molecules HCl and HF \cite{2011A&A...533L...6A}.

To represent an oxygen-rich envelope, we observed the RSG VY Canis Majoris (VY CMa). In contrast to IRC+10216, whose initial stellar mass is $3-5$ $M_{\Sun}$ \citep{2015MNRAS.449..220M}, VY CMa has a mass of 17 M$_{\Sun}$, and is ejecting material at a higher rate of about $6\times10^{-4}$ M$_{\Sun}$ yr$^{-1}$ \citep{Shenoy:2016AJ}. Due to this dramatic mass loss, the envelope of VY CMa is highly asymmetric, consisting of various dusty knots and outflows with non-uniform velocity structure \citep{Smith:2001A}. Because most of the carbon in this envelope resides in CO, VY CMa does not show the same diversity of C-bearing species seen in carbon-rich sources like IRC+10216. It instead exhibits oxygen-driven chemical structure most notably marked by various metal and non-metal oxides, as well as strong OH maser emission \citep{2007Natur.447.1094Z,2018ApJ...856..169Z}. 

In this paper, we report our search for SiH, PH, and FeH toward these evolved stars. In Section \ref{observations}, we summarize our observational strategies and specifications. In Section \ref{Results}, we state our findings and numerically constrain the abundances of these molecules by adopting excitation temperatures and source sizes characteristic of molecules formed in the inner regions of CSEs. Finally, in Section \ref{discussion} we discuss how our results relate to the current theoretical framework for chemical processes in these circumstellar environments, and outline prospects for future related works.

\section{Observations}
\label{observations}

\begin{deluxetable*}{lcccc}

    \tablecaption{Frequency bands observed with SOFIA GREAT}
    \tablewidth{\textwidth}
    \tablehead{\colhead{Targeted Molecule} & \colhead{Rest Band (MHz)} & \colhead{Image Band (MHz)} & Channel Width (\kms)& \colhead{RMS Noise (mK)}}
    \startdata
    \ce{SiH} $\left(3/2-1/2\right)$ & $622220-626220$ & $634220-638220$ & 1.051 & 30.0 \\
    \ce{SiH} $\left(5/2-3/2\right)$ & $1041620-1045620$ & $1053620-1057620$ & 1.051 & 80.1\\
    \ce{FeH} & $1322373-1324996$ & $1325190-1327825$ & 0.995 & 32.0\\
    \ce{PH} & $1505500-1508130$ & $1508340-1510970$ & 1.019 & 96.7\\
    \enddata
    \tablecomments{RMS noise levels are measured from IRC+10216 observations. Integration time on VY CMa was about half that for IRC+10216, so RMS noise is ${\sim}1.4$ times those listed here. PH was only targeted during observations of IRC+10216. }
\label{tab:obs_windows}
\end{deluxetable*}

Observations of IRC+10216 and VY Canis Majoris were taken with the SOFIA telescope using the GREAT receiver in its 4GREAT configuration, which simultaneously collects spectra from four co-aligned pixels operating between 490--635 GHz, 890--1100 GHz, 1200--1500 GHz, and 2490--2590 GHz, respectively. The spectral windows relevant for this work are shown in Table \ref{tab:obs_windows}. PH was only targeted during observations of IRC+10216, whereas the SiH and FeH windows were utilized for both sources.

Observations were taken on 14 and 20 December, 2018. The pointing positions for the observations of IRC+10216 and VY CMa were $\alpha_\text{J2000}=\hms{9}{47}{57.406}$, $\delta_\text{J2000}=\dms{+13}{16}{43.56}$ and $\alpha_\text{J2000}=\hms{7}{22}{58.329}$, $\delta_\text{J2000}=\dms{-25}{46}{03.24}$ respectively. We used the single point chopping method for background subtraction, in which the pointing is symmetrically switched between the ON (centered source) and OFF ($90"$ offset in RA) positions at a rate of 2.5 Hz. Spectra were adjusted to systemic velocities of  $v_{\rm{lsr}}$~=~$-$26.0~km~s$^{-1}$ and $v_{\rm{lsr}}$~=~19.5~km~s$^{-1}$ for IRC+10216 and VY CMa, respectively \citep{Gong:2015ks,1992ApJ...394..320S}. The SOFIA GREAT instrument is a heterodyne receiver that gathers data as a dual-sideband spectrum. Therefore, every frequency in the rest band has a corresponding image frequency that can contaminate observed spectral features. While this type of blending does affect some of the molecular emission seen in our spectra, we selected IFs for each spectral window that ensured no bright lines of known molecules in the image band would overlap with the targeted transitions. 

Using the setup described above, we searched for the $J=3/2-1/2$ and $J=5/2-3/2$ transitions of SiH, the $J=4-3$ ($N=3-2$) transition of PH. Table \ref{tab:IRC_all} includes the spectroscopic parameters for these transitions. The two $J$ transitions of SiH are split into triplets which we would not expect to resolve given the known expansion velocities of IRC+10216 and VY CMa. The transitions of PH also exhibit hyperfine splitting, and there are ${\sim}10$ distinct emission features from this molecule centered at 1,507,640 MHz. Both molecules have small dipole moments, 0.087 and 0.396 Debye for SiH and PH, respectively \citep{2005JMoSt.742..215M}, making detection difficult even with high sensitivity. The dipole moment of PH is based on an unpublished \textit{ab initio} calculation by H. S. P. M{\"u}ller.\footnote{\url{cdms.astro.uni-koeln.de}}

Although the rotational spectrum of FeH is challenging both to observe and to assign in the laboratory, a number of lines have been measured and reported by \citet{Brown:2006gy}.  The $\Omega$ = 5/2; $J = 5/2 \rightarrow 7/2$ transition of FeH of the $X^{4}\Delta$ state was selected because it is the most readily observable transition of the molecule with a frequency and intensity observable by the SOFIA telescope. This transition is split into a pair of $\Lambda$-doubled lines at 1316.8387 GHz and 1324.7719 GHz that are predicted to be of similar intensities. While observing with the SOFIA telescope, we decided to focus only on the higher frequency feature so that integration time at that frequency would be maximized.

Spectral data were reduced using the \texttt{CLASS} program as part of the \texttt{GILDAS} software package.\footnote{\url{http://www.iram.fr/IRAMFR/GILDAS/}} Reduction consisted of removing a first-order baseline, calibrating to main beam temperature scale, and smoothing spectra to a resolution of 1.0 \kms.

\begin{deluxetable*}{lccccccccc}[!t]
\tablecolumns{9}
\tablecaption{List of all detected and targeted rotational transitions toward IRC+10216. }
 \tablehead{\colhead{Molecule} & \colhead{Transition} & \colhead{Frequency} & \colhead{$\int TdV$}  & \colhead{$\Delta V$} & \colhead{$E_{up}$}& \colhead{$S_{ij}\mu^2$} & \colhead{$\theta_{B}$} & \colhead{Comments}\\
  & & (MHz) & (K $\cdot$ \kms) & (\kms) & (K) & (D$^2$) & (arcsec) &  }
 \startdata
 $^{\dagger}$CS & $J=13 - 12$ & \phn636 531.5 $\pm$0.13 & 16.0 $\pm$ 0.1 & 18.1 $\pm$ 0.1 & 213.9 & 49.68 & 42.2 &  \\ \\\hline 
 $^{\dagger}$SiS$_{\nu=0}$ & $J=35 - 34$ & \phn634 398.7 $\pm$0.33 & 4.27 $\pm$ 0.08 & 17.8 $\pm$ 0.4 & 548.5 & 105.4 & 42.3 &  \\ \\
 $^{\dagger}$SiS$_{\nu=1}$ & $J=58 - 57$ & 1043 196.3 $\pm$13\phd\phn & 1.40 $\pm$ 1.0 & 9.92 $\pm$ 8.7 & 2551 & 173.6 & 25.8 &  \\
 $^{\dagger}$SiS$_{\nu=2}$ & $J=59 - 58$ & 1055 811.2 $\pm$25\phd\phn & 0.297 $\pm$ 0.67 & 5.07 $\pm$ 13 & 3659 & 177.8 & 25.4 &  \\ \\
 $^{\dagger}$Si$^{34}$S$_{\nu=0}$ & $J=60 - 59$ & 1054 579.5 $\pm$9.9\phn & 2.66 $\pm$ 1.3 & 18.0 $\pm$ 6.6 & 1547 & 180.7 & 25.5 &  \\ \\\hline
 $^{\dagger}$ $^{30}$SiO$_{\nu=0}$ & $J=15 - 14$ & \phn635 221.2 $\pm$8.6\phn & 0.615 $\pm$ 0.24 & 20.5 $\pm$ 8.2 & 244.0 & 144.0 & 42.3 &  \\ \\\hline
 $^{\dagger}$SiH & $J=3/2-1/2$ & \phn624 924.7$^{b}$\phd\phn\phd\phn\phn\phn & $<0.172$ & -- & 29.99 & 0.0102$^a$ & 43.0 & \\ \\
 $^{\dagger}$SiH & $J=5/2-3/2$ & 1043 917.9$^{b}$\phd\phn\phd\phn\phn\phn & $<0.313$ & -- & 80.09 & 0.0234$^a$ & 25.7 &  \\ \\\hline
 $^{\dagger}$PH & $N=3-2,J=4-3$ & 1507 639.6$^{b}$\phd\phn\phd\phn\phn\phn & $<1.410$ & -- & 144.2 & 2.346$^a$ & 17.8 & \\ \\\hline
 \phn FeH & $\Omega=5/2,J=5/2-7/2$ & 1324 771.4$^{b}$\phd\phn\phd\phn\phn\phn & $<0.449$ & -- & 297.0 & 11.3 & 20.4 & \\ \\\hline
 $^{\dagger}$HNC$_{\nu=0}$ & $J=7-6$ & \phn634 510.8$^{b}$\phd\phn\phd\phn\phn\phn & ? & ? & 121.8 & 65.12 & 42.3 &  (1) \\  \\\hline 
 $^{\dagger}$HCN$_{\nu_2=2}$ & $J=7-6,l=2f$ & \phn623 142.9 $\pm$0.90 & 0.612 $\pm$ 0.07 & 7.04 $\pm$ 0.8 & 2172 & 162.1 & 43.1 &  \\ 
 $^{\dagger}$HCN$_{\nu_2=2}$ & $J=7-6,l=2e$ & \phn623 306.4$^{b}$\phd\phn\phd\phn\phn\phn & ? & ? & 2172 & 162.1 & 43.1 & (2) \\ 
 $^{\dagger}$HCN$_{\nu_2=2}$ & $J=7-6,l=0$ & \phn623 336.6$^{b}$\phd\phn\phd\phn\phn\phn & ? & ? & 2150 & 176.4 & 43.1 & (2) \\
 $^{\dagger}$HCN$_{\nu_2=1}$ & $J=7-6,l=1f$ & \phn623 359.2 $\pm$0.33 & 10.3 $\pm$ 0.1 & 21.4 $\pm$ 0.3 & 1144 & 178.1 & 43.1& (2) \\ \\
 $^{\dagger}$HCN$_{\nu_3=1}$ & $J=12-11$ & 1055 724.5 $\pm$5.4\phn & 6.66 $\pm$ 1.2 & 16.5 $\pm$ 3.6 & 3346 & 319.8 & 25.4 & \\
 $^{\dagger}$HCN$_{\nu_1=1}$ & $J=12-11$ & 1055 501.8 $\pm$11\phd\phn & 3.31 $\pm$ 0.81 & 25.4 $\pm$ 5.3 & 5094 & 327.7 & 25.4 & \\ \\
 $^{\dagger}$HCN$_{\nu_2=2}$ & $J=17-16,l=0$ & 1510 521.2 $\pm$2.7\phn & 21.1 $\pm$ 1.4 & 15.9 $\pm$ 1.1 & 2684 & 428.4 & 17.8 &  \\ \\
 $^{\dagger}$HCN$_{\nu_2=3}$ & $J=17-16,l=1e$ & 1507 906.9 $\pm$3.4\phn & 3.62 $\pm$ 0.49 & 9.43 $\pm$ 1.4 & 2684 & 414.5 & 17.8 &  \\ \\\hline
\enddata
\tablecomments{ \textbf{ (1)} H$^{35}$Cl $J=1-0$ is blended with HNC $J=7-6$ from the image band (see Fig. \ref{fig:IRC_SiH624}). Degree of contamination is unclear, but agreement with \cite{2014ApJ...790L..27A} suggests H$^{35}$Cl is the dominant contributor of the feature. \textbf{(2)} Three transitions of vibrationally excited \ce{HCN} blend around 623350 MHz; HCN$_{\nu_2=1}$ is the brightest component due to its lower energy, therefore our Gaussian fit was attributed to it. \textbf{(3)} $^{41}$KCl blended with SiC$_2$ $29_{2,28}-28_{2,27}$ from the image band. \textbf{(4)} SiC$_2$ and its isotopologue $^{29}$SiC$_2$ blend around 1055000 MHz. \textbf{(5)} Vibrationally excited K$^{37}$Cl transitions are seen at the edge of our 1 THz band. Their emission is blended and difficult to analyze due to the frequency cutoff. \textbf{(6)} Unidentified lines can be seen in Figure \ref{fig:IRC_SiH1043}; we do not report line areas or widths for these features as they either had too low S/N to obtain a reliable fit, or exhibit an anomalous line profile that is not well-modeled by a Gaussian.\\
$^a$Sum of unresolved hyperfine components\\
$^{b}$No fit performed, literature frequency listed \\
$^{c}$Unclear whether feature is present in rest or image band; therefore, we report the rest frequency as part of the molecule name, and list the corresponding image frequency in the third column. \\
$^{\dagger}$Spectroscopic data taken from the CDMS catalogue \citep{2005JMoSt.742..215M}\\
$^{\dagger\dagger}$Spectroscopic data taken from the JPL catalogue:    \url{https://spec.jpl.nasa.gov}.}
\label{tab:IRC_all}
\vspace{-20.93102pt}
\end{deluxetable*}

\setcounter{table}{1}
\begin{deluxetable*}{lccccccccc}[!t]
\tablecolumns{9}
\tablecaption{Continued. }
 \tablehead{\colhead{Molecule} & \colhead{Transition} & \colhead{Frequency} & \colhead{$\int TdV$}  & \colhead{$\Delta V$} & \colhead{$E_{up}$}& \colhead{$S_{ij}\mu^2$} & \colhead{$\theta_{B}$} & \colhead{Comments}\\
  & & (MHz) & (K $\cdot$ \kms) & (\kms) & (K) & (D$^2$) & (arcsec) &  }
 \startdata
 $^{\dagger}$SiC$_2$ & $J_{K_a,K_c}=26_{6,20}-25_{6,19}$ & \phn625 589.0 $\pm$3.9\phn & 0.664 $\pm$ 0.12 & 23.5 $\pm$ 5.3 & 470.6 & 141.0 & 42.9 &  \\ 
 $^{\dagger}$SiC$_2$ & $J_{K_a,K_c}=27_{4,24}-26_{4,23}$ & \phn634 764.8 $\pm$4.3\phn & 0.940 $\pm$ 0.14 & 28.0 $\pm$ 4.8 & 461.7 & 150.5 & 42.3 &  \\ 
 $^{\dagger}$SiC$_2$ & $J_{K_a,K_c}=30_{0,30}-29_{0,29}$ & \phn636 350.9 $\pm$3.6\phn & 0.503 $\pm$ 0.12 & 12.1 $\pm$ 4.0 & 480.8 & 171.1 & 42.2 &  \\ 
 $^{\dagger}$SiC$_2$ & $J_{K_a,K_c}=29_{2,28}-28_{2,27}$ & \phn636 820.3$^{b}$\phd\phn\phd\phn\phn\phn & ? & ? & 478.1 & 164.1 & 42.2 & (3) \\  \\
 $^{\dagger}$SiC$_2$ & $J_{K_a,K_c}=43_{8,35}-42_{8,32}$ & 1054 995.3$^{b}$\phd\phn\phd\phn\phn\phn & ? & ? & 1208 & 238.1 & 25.5 & (4) \\  \\
 $^{\dagger29}$SiC$_2$ & $J_{K_a,K_c}=44_{6,38}-43_{6,37}$ & 1054 973.3$^{b}$\phd\phn\phd\phn\phn\phn & ? & ? & 1208 & 245.8 & 25.5 & (4) \\  \\\hline
 $^{\dagger\dagger}$H$^{35}$Cl & $J=1-0$ & \phn625 922.7 $\pm$0.76  & 3.67 $\pm$ 0.10 & 26.2 $\pm$ 0.83 & 30.04 & 2.458 & 42.9 & (1)\\ \\
 $^{\dagger\dagger}$H$^{37}$Cl & $J=1-0$ & \phn624 972.4 $\pm$7.3\phn & 0.932 $\pm$ 0.29 & 23.3 $\pm$ 8.3 & 29.99 & 2.458 & 43.0 &  \\ \\\hline
 $^{\dagger}$K$^{37}$Cl$_{\nu=0}$ & $J=86-85$ & \phn634 623.9 $\pm$2.7\phn & 0.663 $\pm$ 0.08 & 18.1 $\pm$ 2.7 & 1333 & 9069 & 42.3 & \\ \\
 $^{\dagger}$K$^{37}$Cl$_{\nu=0}$ & $J=145-144$ & 1045 482.5 $\pm$20\phd\phn & 1.35 $\pm$ 1.31 & 18.0 $\pm$ 13 & 3729 & 15297 & 25.7 & \\ 
 $^{\dagger}$K$^{37}$Cl$_{\nu=1}$ & $J=146-145$ & 1045 577.8$^{b}$\phd\phn\phd\phn\phn\phn & ? & ?  & 4150 & 15581 & 25.7 & (5)\\
 $^{\dagger}$K$^{37}$Cl$_{\nu=2}$ & $J=147-146$ & 1045 596.7$^{b}$\phd\phn\phd\phn\phn\phn & ? & ? & 4567 & 15871 & 25.7 & (5) \\ \\
 $^{\dagger}$ $^{41}$KCl & $J=84-83$ & \phn623 616.1$^{b}$\phd\phn\phd\phn\phn\phn & ? & ? & 1279 & 8858 & 43.1 & (3) \\ \\\hline
 $^{\dagger}$AlCl & $J=43-42$ & \phn624 431.2 $\pm$3.4\phn & 0.630 $\pm$ 0.19& 10.8 $\pm$ 4.2 & 660.5 & 45.96$^*$ & 43.0 &  \\ \\\hline
 U1042852 & -- & 1056 383$^c$\phd\phn\phn\phd\phn\phn\phn & -- & -- & -- & -- & 25.8/25.4 & (6) \\
 U1042987 & -- & 1056 248$^c$\phd\phn\phn\phd\phn\phn\phn & -- & -- & -- & -- & 25.8/25.4 & (6) \\
 U1044507 & -- & 1054 728$^c$\phd\phn\phn\phd\phn\phn\phn & -- & -- & -- & -- & 25.7/25.5 & (6) \\
\enddata

\label{tab:IRC_all}
\end{deluxetable*}

\begin{figure*}
    \centering
    \includegraphics[width=0.85\textwidth]{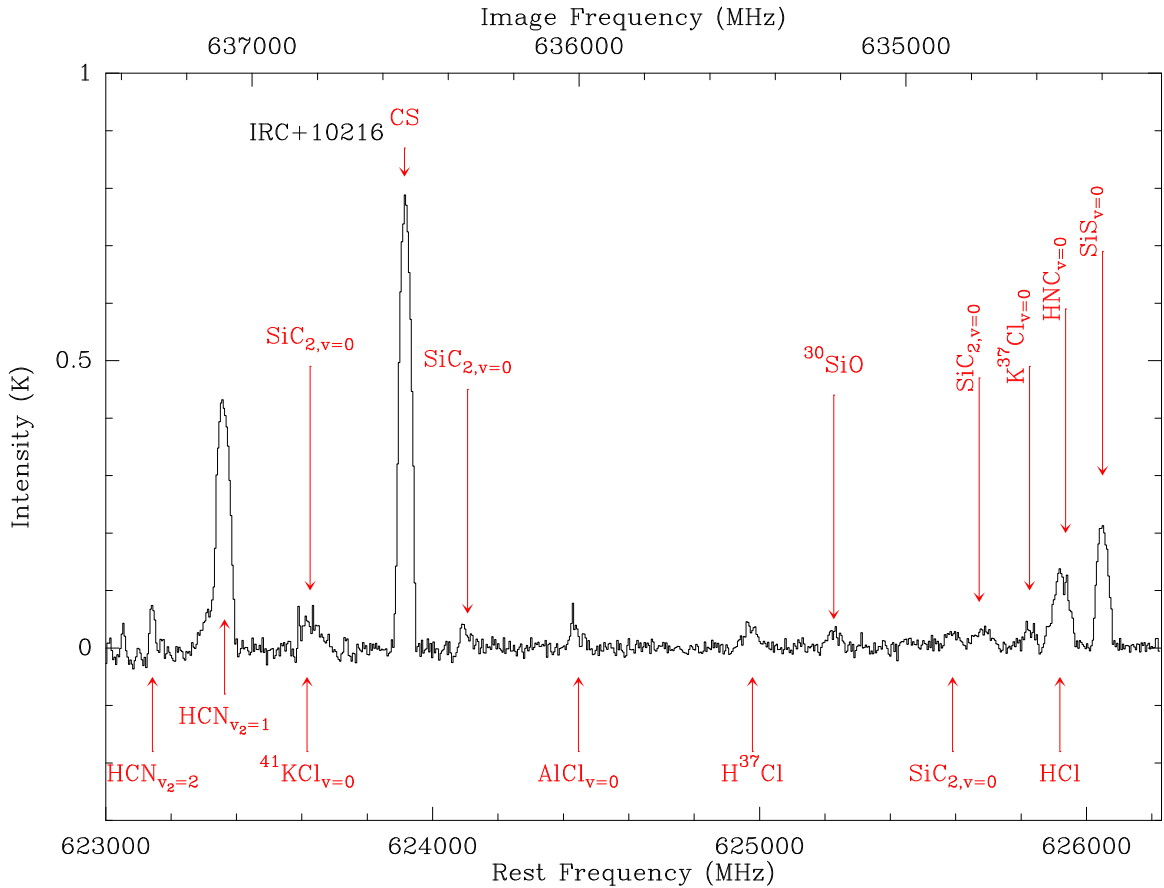}
    \caption {Continuum subtracted SOFIA GREAT spectrum of IRC+10216 around targeted 624925 MHz transition of SiH. Emission lines contributed from the rest frequency band are written near the lower x-axis, and those contributed from the image band are written near the upper x-axis. }  
    \label{fig:IRC_SiH624}
\end{figure*}

\begin{figure*}
    \centering
    \includegraphics[width=0.85\textwidth]{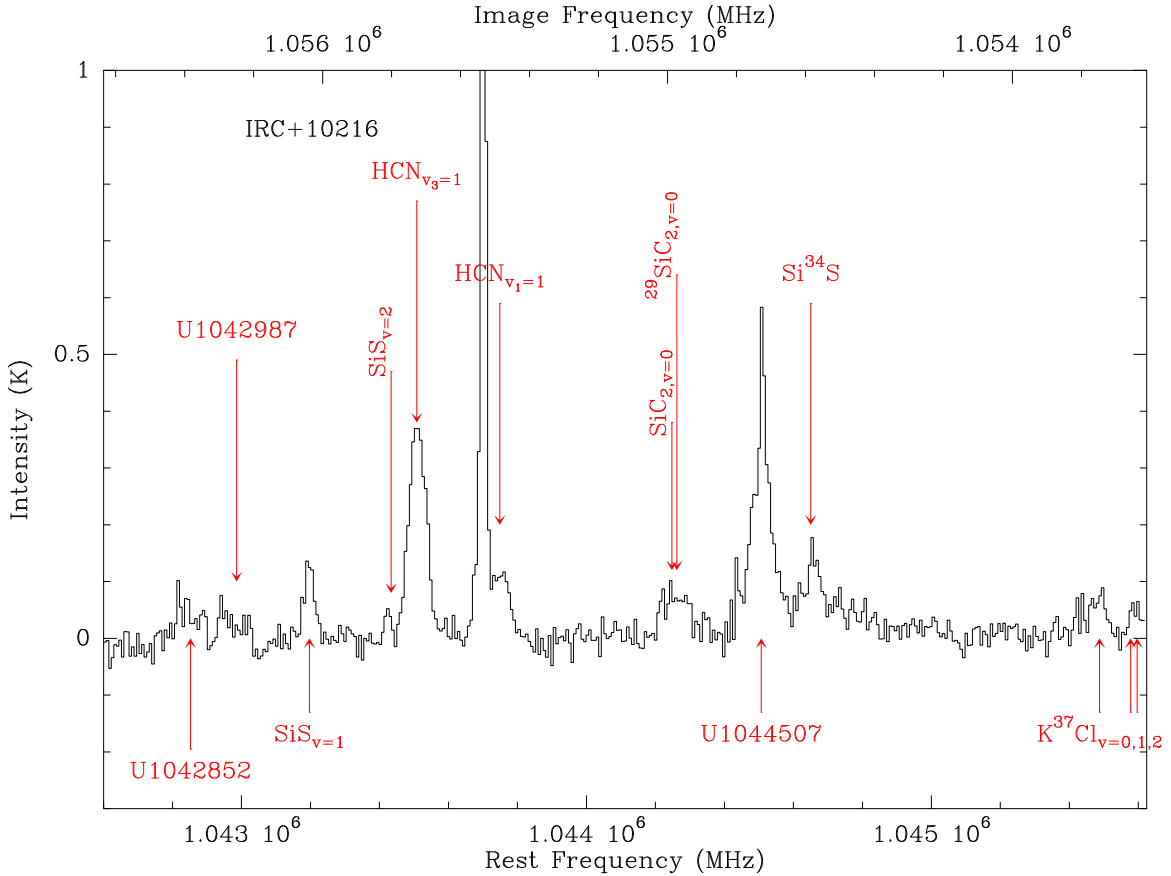}
    \caption {Continuum subtracted SOFIA GREAT spectrum of IRC+10216 around targeted 1043918 MHz transition of SiH. Emission lines contributed from the rest frequency band are written near the lower x-axis, and those contributed from the image band are written near the upper x-axis. Unidentified lines are labeled with their observed frequency in the rest band even though they could be contributed from the image band. Corresponding image frequencies for each U line are shown in Table \ref{tab:IRC_all}.}  
    \label{fig:IRC_SiH1043}
\end{figure*}

\section{Results \& Analysis}
\label{Results}

\begin{figure*}
\hspace{0.6in}
\subfloat[SiH $J=3/2-1/2$]{%
  \includegraphics[height=2.75in]{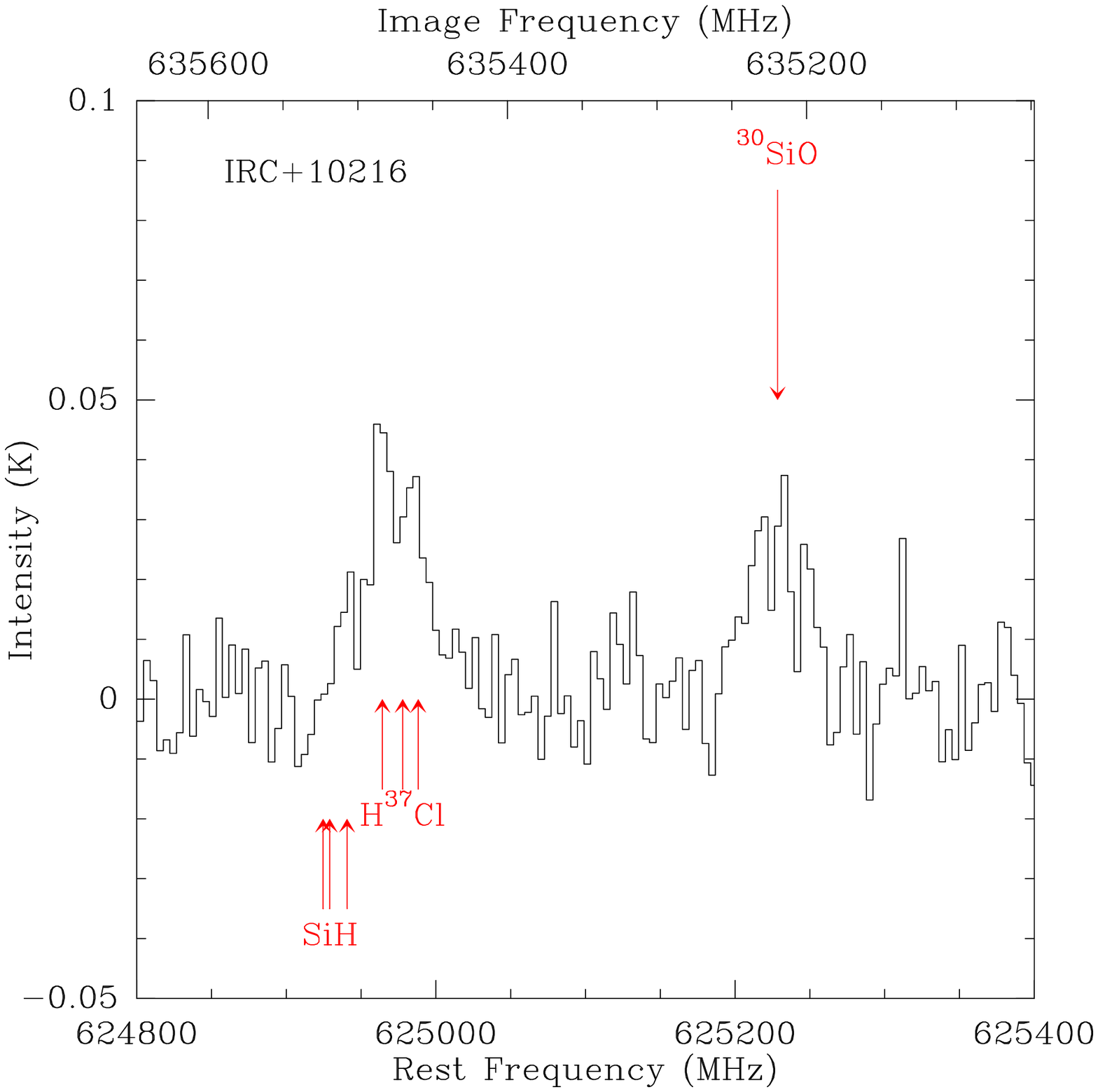}%
  \label{plot:IRC_SiH_low_zoom}%
}\hspace{0.3in}
\subfloat[SiH $J=5/2-3/2$]{%
  \includegraphics[height=2.75in]{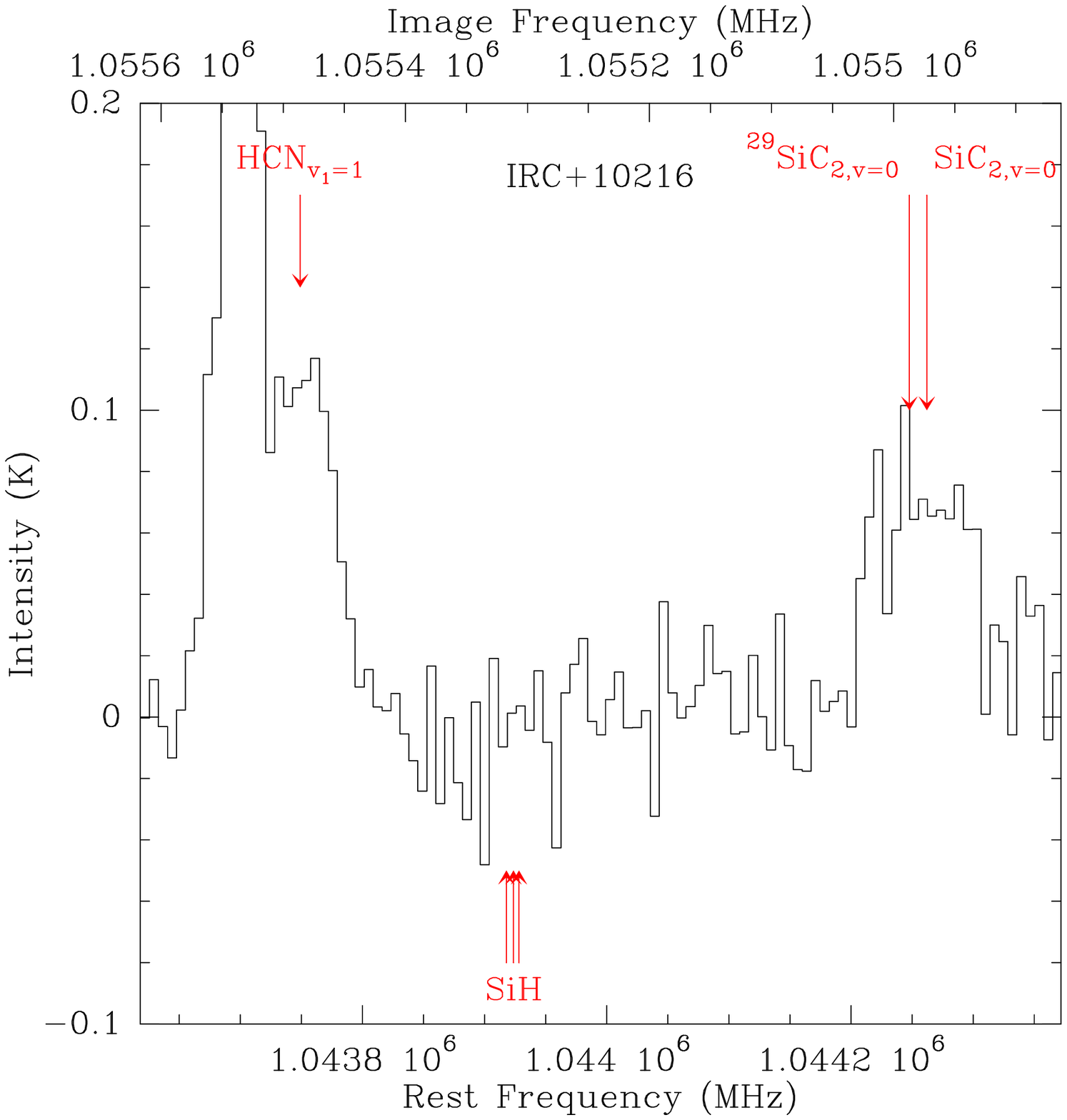}%
  \label{plot:IRC_SiH_high_zoom}%
}

\hspace{0.6in}\subfloat[FeH $\Omega=5/2,J=5/2-7/2$]{%
  \includegraphics[height=2.75in]{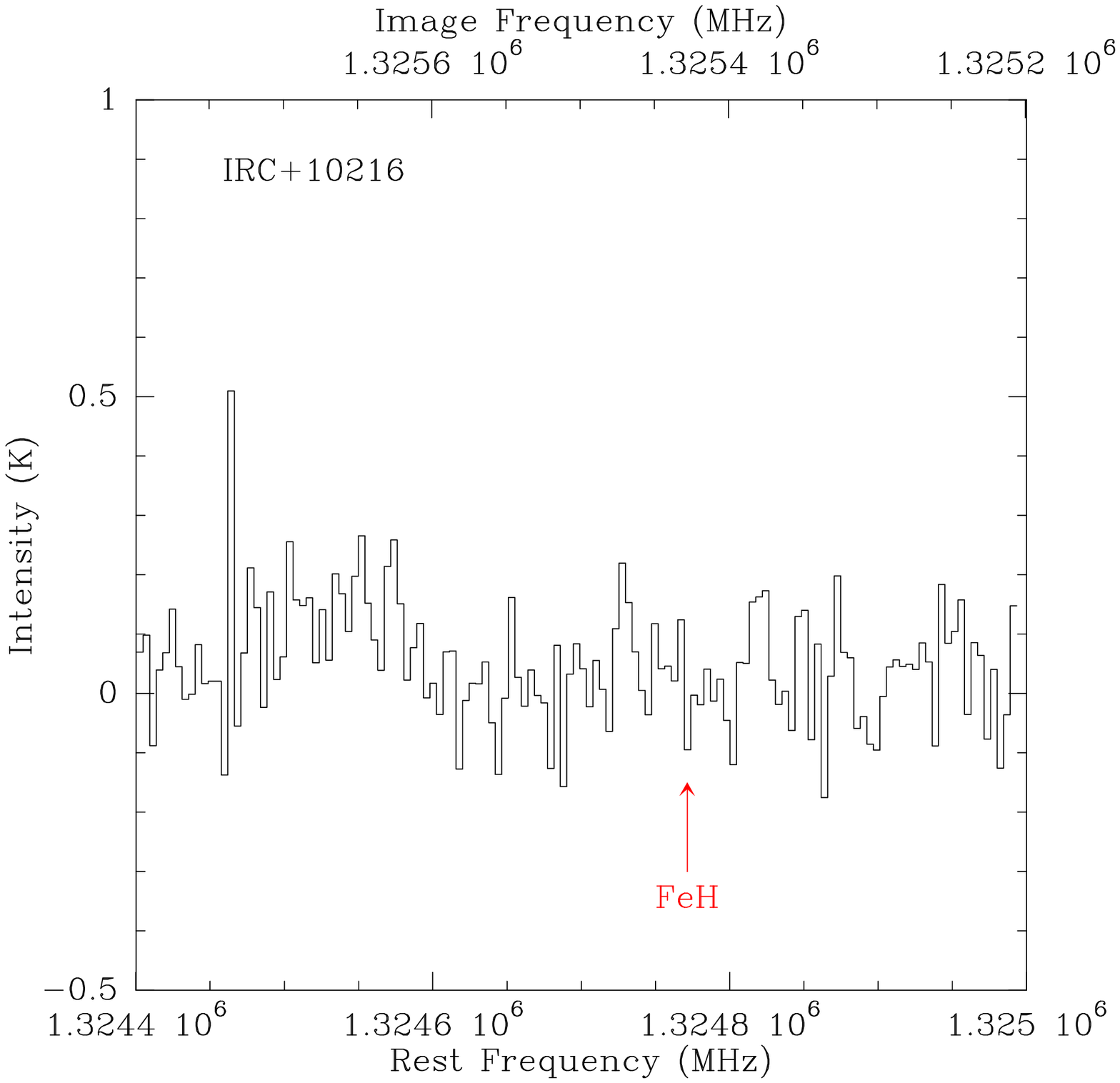}%
  \label{plot:IRC_FeH_zoom}%
}\hspace{0.29in}
\subfloat[PH $N=3-2$,$J=4-3$]{%
  \includegraphics[height=2.75in]{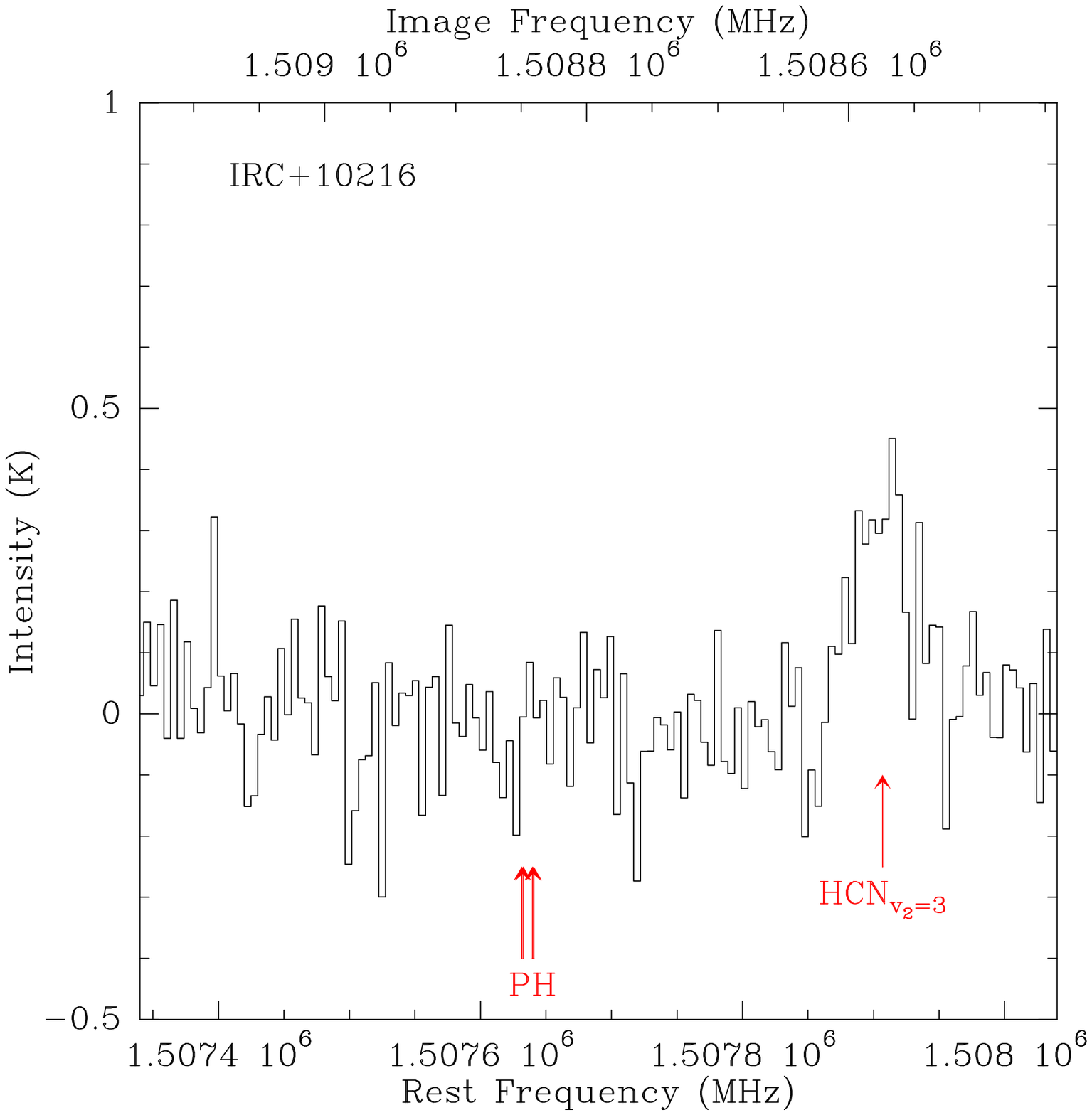}%
  \label{plot:IRC_PH_zoom}%
  }
\caption{SOFIA GREAT spectra of IRC+10216 showing targeted emission lines of SiH $J=3/2-1/2$ (top left), SiH $J=5/2-3/2$ (top right), FeH $\Omega=5/2,J=5/2-7/2$ (bottom left), and PH $J=4-3$ (bottom right).}
\label{fig:IRCzooms}
\end{figure*}

\subsection{IRC+10216}
In our search for these hydrides, we detected a total of 26 transitions which we assign to 11 different species toward IRC+10216. Figures \ref{fig:IRC_SiH624} and \ref{fig:IRC_SiH1043} show the full calibrated and baseline-subtracted spectra for our 624 and 1048 GHz windows. At these frequencies, emission from vibrationally excited HCN dominates most of the observed bands and accounts for over a quarter of the observed transitions. All other molecules have been previously observed in IRC+10216, with the notable exception of three unidentified features in our 1.04 THz spectral window (labeled in Fig.~\ref{fig:IRC_SiH1043}).  A list of detected transitions is shown in Table \ref{tab:IRC_all}. 
Each spectral line that was not severely affected by blending was fit with a Gaussian profile in \texttt{CLASS}. Resulting profile widths and integrated line temperatures are included in \ref{tab:IRC_all} along with their fitting errors.

We compute column densities and abundances for the observed rotational transitions following the formalism of \citet{Hollis:2004uh} for molecular emission described by a single excitation temperature (Eq. \ref{eq:Hollis_col}). $T_{ex}$ is the excitation temperature (K); $T_{bg}$ is the background continuum temperature and assumed to be 2.7~K; $k$ is Boltzmann's constant (J~K$^{-1}$); $h$ is Planck's constant (J~s); $Q$ is the rotational partition function at $T_{ex}$; $E_{u}$ is the upper state energy of the transition (K); $\int TdV$ is the velocity-integrated line area (K $\cdot$ \kms) which can be expressed as the product of the peak antenna temperature and the line full width at half maximum; $S_{ij}\mu^2$ is the transition line strength (Debye$^2$); $B$ is the beam filling factor; and $\eta_B$ is the beam efficiency.
\begin{equation}
N_T=  \frac{1}{2}  \frac{3k}{8\pi^3}  \sqrt{\frac{\pi}{\ln 2}}\frac{Qe^\frac{E_u}{T_{ex}}\int TdV}{B\nu S_{ij}\mu^2\eta_B} \frac{1}{1-\frac{e^{\frac{h\nu}{kT_{ex}}{-1}}}{e^{\frac{h\nu}{kT_{bg}}{-1}}}}
\label{eq:Hollis_col}
\end{equation}

For most molecules, including \ce{SiH} and \ce{PH}, spectroscopic constants $Q$ and $S_{ij}\mu^2$ were obtained from the CDMS and JPL\footnote{https://spec.jpl.nasa.gov} molecular databases \citep{2005JMoSt.742..215M}. The only exception to this was FeH, whose partition function was calculated and published by \cite{2016A&A...588A..96B}. All partition functions were interpolated to our assumed rotational temperatures using a quartic fit on the measured $Q$ values. 

Unfortunately, there is a lack of spectroscopic data concerning the line strengths of FeH transitions in the $X^{4}\Delta$ state. Because of this, we approximate the $S_{ij}$ value for this molecule as the 2$_1$-1$_1$ transition of a symmetric top \citep{Gordy:1984uy}, yielding a value of $\sim$0.67. The dipole moment was calculated to be 4.1 Debye using Gaussian 09 \citep{gaussian09B1} at the CCSD(T)/aug-cc-pVQZ level of theory and basis set. From this, we obtain a total $S_{ij}\mu ^2$ value of 11.3 Debye$^2$ (Tables~\ref{tab:IRC_all},\ref{tab:VY_all}). We note that while the above approximation may be a simplified treatment of this molecule, it is still a sufficient choice given the intent of our search, and the current body of spectroscopic data for FeH and similar molecules. We are confident that this value yields column densities and abundances of FeH to within an order of magnitude.

To derive fractional abundances (relative to \ce{H2}) for species detected toward IRC+10216, we employ the formula from \cite{Gong:2015ks} that gives us the average \ce{H2} column density over a radius $R$ in a spherically expanding wind with velocity $V_{exp}$, mass-loss rate $\dot{M}$, and mean molecular weight $\mu$:
\begin{equation}
    N_{H_2}=\frac{\dot{M}}{\pi RV_{exp}\mu m_H}
    \label{eq:H2col}
\end{equation}
For IRC+10216, we take $\mu=2.3$ amu, $V_{exp}=14.5$ \kms, and $\dot{M}=2.0\times10^{-5}$ M$_{\Sun}$yr$^{-1}$ from \cite{Massalkhi:611A29M}.

For every molecule, we assume an emission size characteristic of the inner envelope (denoted by the angular radius $\theta_{em}$), and use this to calculate the beam filling factor and radius $R$ to compare with the average \ce{H2} column density. 

Using this method, we obtain abundances relative to \ce{H2} for some well-characterized molecules in these sources that show emission in our frequency bands. In IRC+10216, for CS we calculate $f=5\times10^{-7}$ (assuming $T_{rot}=175$ K and $\theta_{em}=1"$); for H$^{35}$Cl: $f=1\times10^{-7}$ ($T_{rot}=150$ K and $\theta_{em}=2"$); for \ce{SiC2}: $f=2\times10^{-7}$ ($T_{rot}=150$ K and $\theta_{em}=2"$). These results are in agreement with previous studies of IRC+10216 to within a factor of ${\sim}$2--3 \citep{Massalkhi:611A29M,2019A&A...628A..62M,2011A&A...533L...6A}. Thus, we are confident that this treatment may be used to obtain accurate constraints on the hydrides targeted in this work.

\begin{figure}[h!]
\hspace{0.4in}\subfloat[SiH $J=3/2-1/2$]{%
  \includegraphics[height=2.38in]{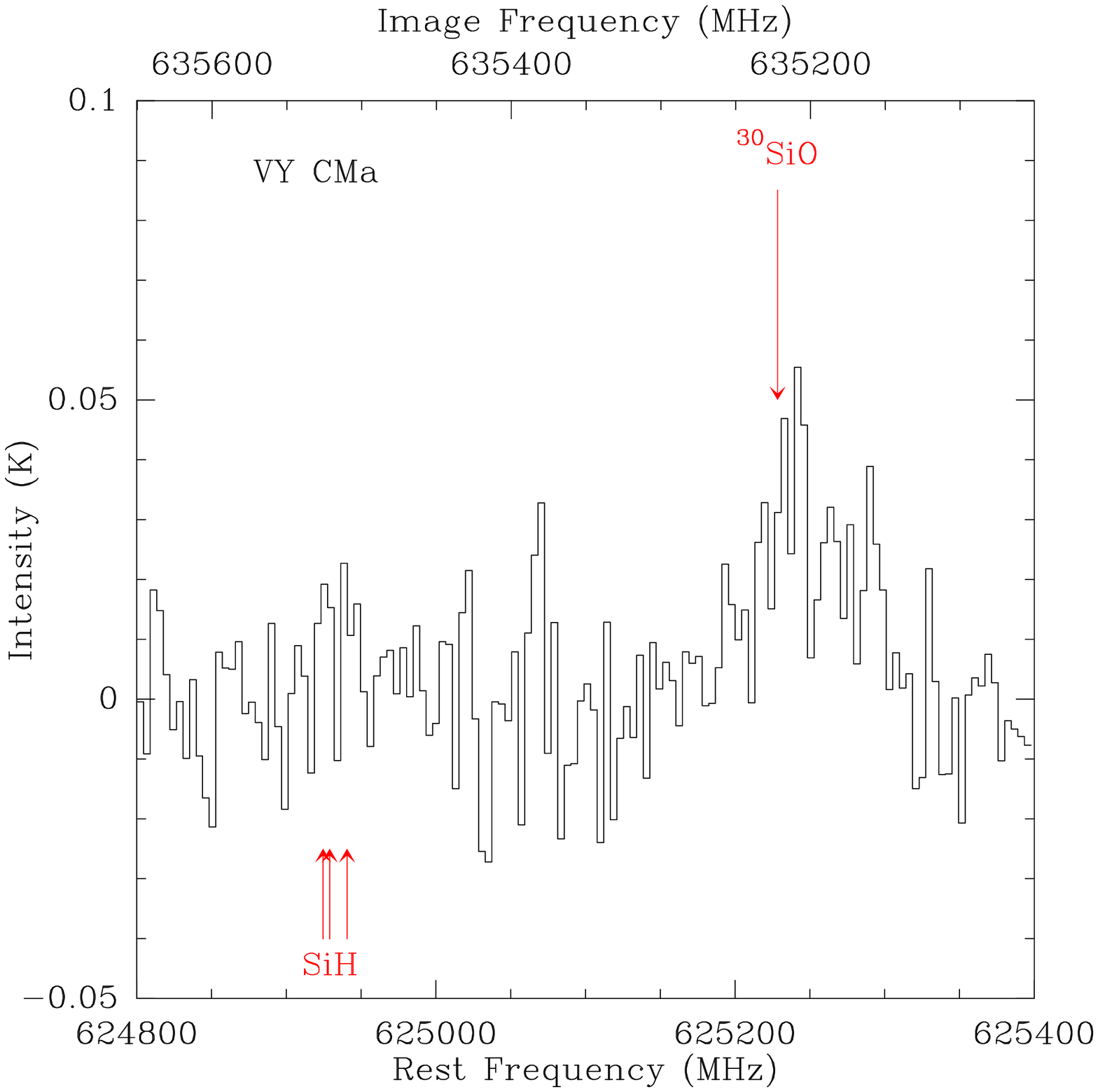}%
  \label{plot:VY_SiH_low_zoom}%
}

\hspace{0.4in}\subfloat[SiH $J=5/2-3/2$]{%
  \includegraphics[height=2.38in]{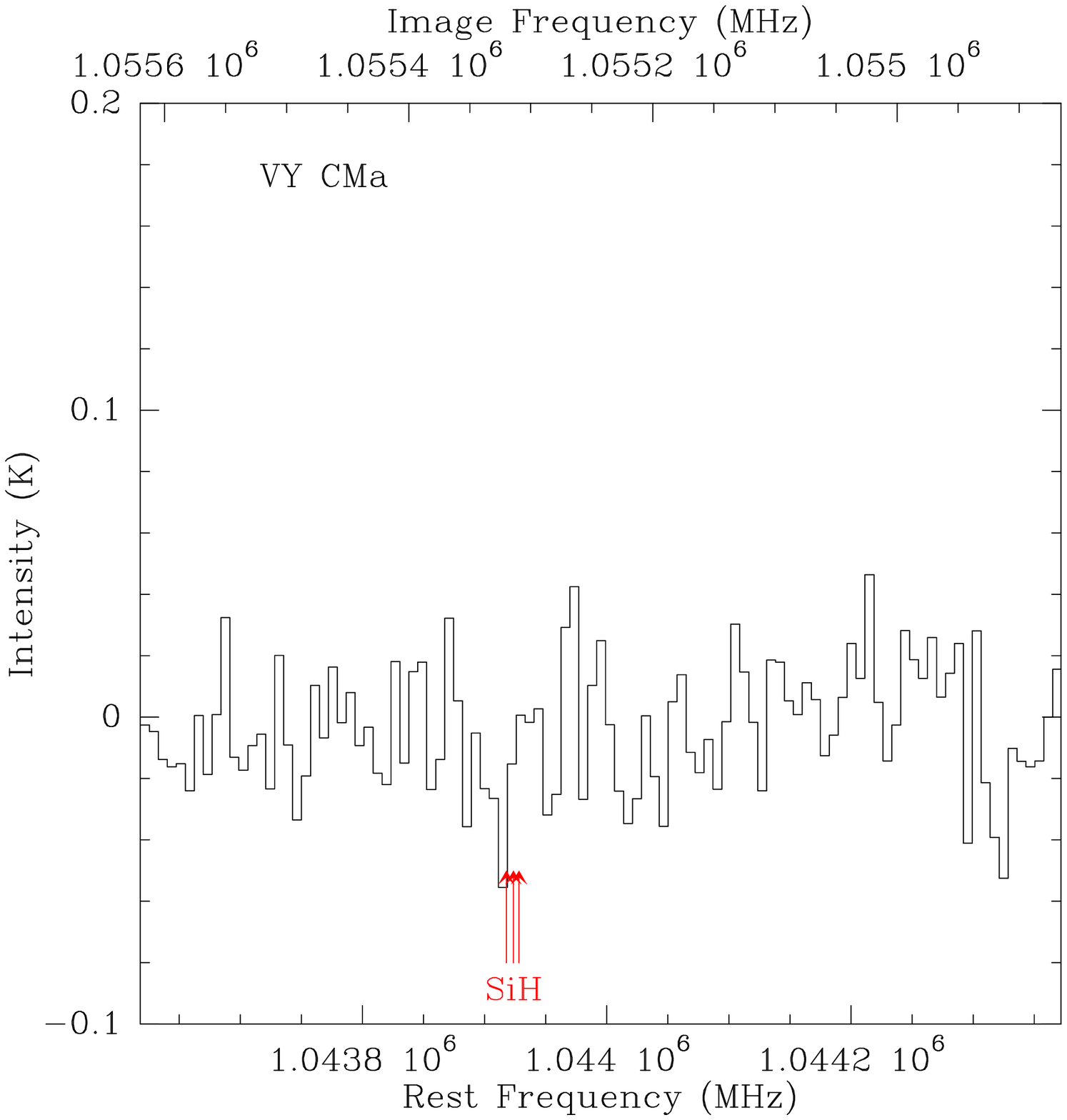}%
  \label{plot:VY_SiH_high_zoom}%
}

\hspace{0.4in}\subfloat[FeH $\Omega=5/2,J=5/2-7/2$]{%
  \includegraphics[height=2.38in]{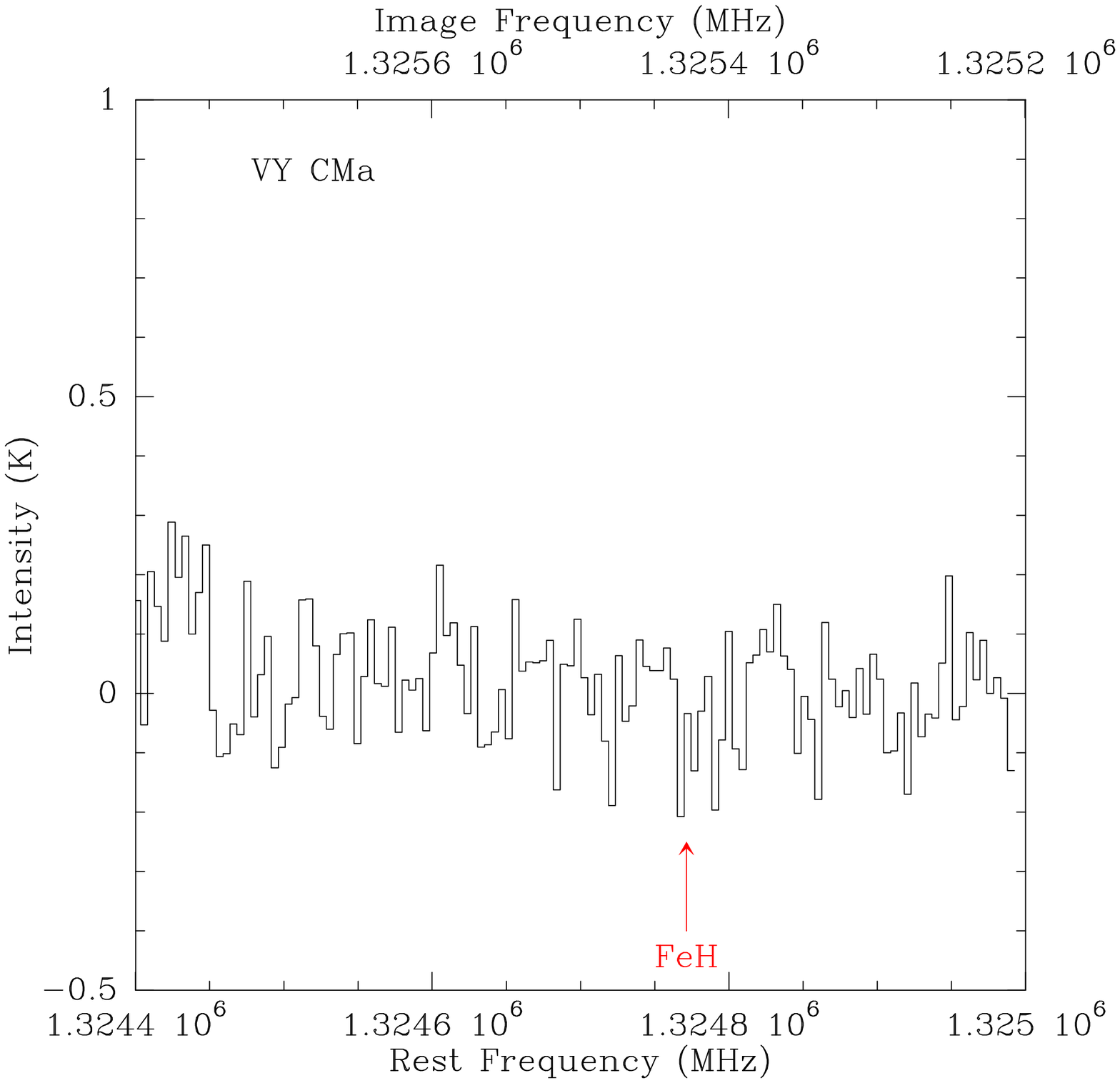}%
  \label{plot:VY_FeH_zoom}%
  }
\caption{SOFIA GREAT spectra of VY CMa showing targeted emission lines of SiH $J=3/2-1/2$ (top row), SiH $J=5/2-3/2$ (middle row), and FeH $\Omega=5/2,J=5/2-7/2$ (bottom row).}
\label{fig:VYzooms}
\end{figure}

Neither SiH, PH, nor FeH were detected in the C-rich envelope of IRC+10216 (see Fig.~\ref{fig:IRCzooms}).  Upper limits on the integrated line fluxes, obtained assuming a width of 14.5 \kms and a peak antenna temperature equal to the 1$\sigma$ rms noise level of the observations, are shown in Table \ref{tab:IRC_all}.

The $J=4-3$ transition of PH has a similar Einstein $A_{ij}$ coefficient to the $1-0$ transition of HCl. Assuming these molecules have similar collisional excitation rates, we expect PH to thermalize at a similar critical density to HCl given by \cite{2011A&A...533L...6A} (${\sim}10^{7}$ \cmnegthree). In IRC+10216, gas densities are below this beyond radii of $1.3\times10^{15}$ cm (0.7"). Because of this, we assume all PH emission is confined to a 2" source size, and has an effective excitation temperature of 250 K in this region, similar to those measured by \cite{2011ApJS..193...17P} for \ce{SiC2} and \ce{HC3N}. We derive an upper limit column density of $N_{\rm{PH}}<7\times10^{14}$~\cmnegtwo~, corresponding to a fractional abundance relative to \ce{H2} of $f_{\rm{PH}}<4\times10^{-8}$.

The $A_{ij}$ values of SiH transitions are much lower than PH, so we expect the excitation of this molecule to be described by LTE conditions out to larger radii than PH and HCl. In this case, we assume a 5" maximum radius of emission and an excitation temperature of 150 K. Given these constraints, the higher temperature $J=5/2-3/2$ transition is expected to produce a stronger signal than the one at 625 GHz. Using the spectroscopic and observational parameters for this line in the calculation described above, we derive an upper limit column density of $N_{\rm{SiH}}<2\times10^{15}$~\cmnegtwo~, and a fractional abundance relative to \ce{H2} of $f_{\rm{SiH}}<3\times10^{-7}$.

Finally, for \ce{FeH}, we assume conditions similar to those of \ce{PH}, adopting an average value of $T_{rot}$~=~250~K and $\theta_{em}=2"$. From this, we derive $N_{\rm{FeH}}<9\times10^{13}$~\cmnegtwo~and $f_{\rm{FeH}}<2\times10^{-7}$.
\vspace{0.2in}
\subsection{VY CMa}
Similar to the case of IRC+10216, we saw no emission from \ce{SiH} nor \ce{FeH} in the O-rich envelope of VY CMa (Fig.~\ref{fig:VYzooms}). Toward this source, we only detect emission from $^{30}$SiO and SiS. Table \ref{tab:VY_all} shows the observational and spectroscopic parameters of these transitions in addition to the targeted lines of SiH and FeH. Much like with IRC+10216, we expect most ro-vibrational emission to come from regions very close to the stellar photosphere (implying compact emission ${\sim}1"$ in radius). This assumption is supported by the fact that all observed molecules exhibit narrow ($<40$ \kms) emission lines characteristic of regions of the CSE where radiation pressure accelerates ejected material.\\
\indent As noted before, the CSE of VY CMa is much less isotropic than that of IRC+10216. \cite{Ziurys:2007A} demonstrate that three separate velocity structures dictate molecular emission profiles in this envelope: a spherically expanding wind centered at the systemic velocity, a tightly collimated blue flow at $v_{lsr}=-4$ \kms, and a redshifted expansion at $v_{lsr}=48$ \kms. Some molecules (e.g.\ \ce{SO}, \ce{SO2}, \ce{HNC}, \ce{CO},  and \ce{OH} masers) exhibit double- or triple-peaked lines arising from the strong emission in the bipolar outflow \citep{Ziurys:2007A,2010ApJS..190..348T}. However, in a \textit{Herschel} HIFI spectral line survey of VY CMa, \cite{2013A&A...559A..93A} note that typical emission profiles above 600 GHz are dominated by the steady wind component. This has been shown to be especially true for all observed transitions of refractory metal-bearing molecules like NaCl, AlCl, AlO, and TiO \citep{Kaminski:2013gk,2013ApJS..209...38K}. Because of this, we assume that SiH and FeH are only present in the spherical wind to calculate their upper limit abundances. This allows us to employ the same quantitative analysis as we did with IRC+10216, by using equations \ref{eq:Hollis_col} and \ref{eq:H2col}.\\
\indent We adopt a mass-loss rate of $1\times10^{-4}$ M$_\Sun$ yr$^{-1}$, a terminal expansion velocity of 40 \kms, and a mean molecular weight of 2.3 \citep{Ziurys:2007A,2013A&A...549A...6K}. We assume SiH and FeH emission come from within a radius of 1", which is typical of molecules in this source \citep{2013ApJS..209...38K}. We adopt line widths equal to the terminal expansion velocity of $v_{exp}=40$~\kms, and rotational excitation temperatures of 250~K for both SiH and FeH. Given these constraints, we derive an upper limit column density $N_{\rm{SiH}}<2\times10^{17}$~\cmnegtwo~and fractional abundance relative to \ce{H2}, $f_{\rm{SiH}}<3\times10^{-5}$ in the envelope of VY CMa. And for \ce{FeH} we derive $N_{\rm{FeH}}<2\times10^{14}$~\cmnegtwo~and $f_{\rm{FeH}}<3\times10^{-8}$.\\
\indent It is important to note that we cannot rule out the possibility of these hydrides being present in the bipolar outflow of VY CMa; especially in the case of SiH, as the targeted transitions of this molecule involve lower rotational energy states relative to those of FeH and the species observed in \cite{2013A&A...559A..93A}. However, given our aim of investigating the potential formation and roles in dust grain chemistry of hydrides in CSEs, we find it important to calculate abundances that can be compared with other species known to exist in the \textit{interior} regions of this source (where grain production takes place and the red- and blue-shifted flows are not yet present). In any case, the upper limit abundances w.r.t.\ \ce{H2} reported here are accurate descriptions of the chemical inventory in the spherical component of VY CMa, since an increase of emission in the bipolar flow would imply a lower abundance in the stellar wind.

\begin{deluxetable*}{lccccccccc}
\tablecolumns{9}
\tablecaption{List of all detected and targeted rotational transitions toward VY CMa. }
 \tablehead{\colhead{Molecule} & \colhead{Transition} & \colhead{Frequency} & \colhead{$\int TdV$}  & \colhead{$\Delta V$} & \colhead{$E_{up}$}& \colhead{$S_{ij}\mu^2$} & \colhead{$\theta_{B}$} & \colhead{Comments}\\
  & & (MHz) & (K $\cdot$ \kms) & (\kms) & (K) & (D$^2$) & (arcsec) &  }
 \startdata
 $^{\dagger}$SiS$_{\nu=0}$ & $J=35-34$ & \phn634 377.6 $\pm$9.6\phn & 0.613 $\pm$ 0.19 & 26.0 $\pm$ 7.7 & 548.5 & 105.4 & 42.3 &  \\ \\
 $^{\dagger}$ $^{30}$SiO$_{\nu=0}$ & $J=15-14$ & \phn635 201.4 $\pm$7.6\phn & 1.32 $\pm$ 0.24 & 38.2 $\pm$ 7.6 & 244.0 & 144.0 & 42.3 &  \\ \\\hline
 $^{\dagger}$SiH & $J=3/2-1/2$ & \phn624 924.7$^{b}$\phd\phn\phd\phn\phn & $<0.413$ & -- & 29.99 & 0.0102$^a$ & 43.0 & \\ \\
 $^{\dagger}$SiH & $J=5/2-3/2$ & 1043 917.9$^{b}$\phd\phn\phd\phn\phn & $<0.903$ & -- & 80.09 & 0.0234$^a$ & 25.7 &  \\ \\\hline
 \phn FeH & $\Omega=5/2,J=5/2-7/2$ & 1316 838.7$^{b}$\phd\phn\phd\phn\phn & $<0.960$ & -- & 297.0 & 11.3 & 20.4 & \\ \\\hline
\enddata
\tablecomments{$^a$Sum of unresolved hyperfine components\\
$^{b}$No fit performed, literature frequency listed \\
$^{\dagger}$Spectroscopic data taken from the CDMS catalogue \citep{2005JMoSt.742..215M}.}
\label{tab:VY_all}
\end{deluxetable*}
\section{Discussion}
\label{discussion}
The non detections reported here could be simply due to the weak permanent dipoles of the targeted species, as well as effects related to beam dilution. Nevertheless, additional factors related to the underlying reactivity of these species cannot be ruled out. In this section, we discuss how these non-detections are similarly consistent with previous findings related to SiH, PH, FeH and related molecules and may, in fact, help shed light on the chemistry of CSEs.

\subsection{Iron Chemistry in CSEs}
Due to its refractory nature, a significant fraction of the iron in circumstellar winds is likely incorporated into dust grains \citep{sofia_abundant_1994}. After condensing in the inner regions of the envelope, it can be reintroduced into the gas via non-thermal energetic mechanisms like shocks, which drive grain erosion via mechanisms such as sputtering \citep{tielens_physics_1994,tielens_grain_1996,schilke_SiO_1997,jones_grain_1994,caselli_grain-grain_1997,codella_CHESS_2010}. Nevertheless, \cite{Mauron:2010dk} observed absorption from many refractory atomic elements (e.g. Fe, Cr, Ca) in the gas phase toward a background star near IRC+10216, so one cannot safely assume that depletion of Fe onto grains is the sole underlying cause of the non-detection reported here.

Whether sputtered from grains or by some other mechanisms, once in the gas, iron atoms (or \ce{Fe+}) can react to form FeH or FeCN. The detection of the latter species in IRC+10216 is significant, since it remains the only iron-bearing molecule definitively identified in interstellar or circumstellar environments to date \citep{Zack:2011jx} (although \cite{walmsley_detection_2002} have reported a tentative detection of FeO). As an initial guess, it might be assumed that this gas-phase iron chemistry could lead to FeH via one or more of the following reactions: 

\begin{equation}
\mathrm{Fe} + \mathrm{H_2} \rightarrow \mathrm{FeH} + \mathrm{H}
\label{r1}
\end{equation}

\begin{equation}
\mathrm{Fe} + \mathrm{H} \rightarrow \mathrm{FeH} + h\nu
\label{r2}
\end{equation}

\begin{equation}
\mathrm{Fe} + \mathrm{H_3^+} \rightarrow products
\label{r3}
\end{equation}

\begin{equation}
\mathrm{Fe^+} + \mathrm{H_2} \rightarrow products.
\label{r4}
\end{equation}

\cite{Cherchneff:2012ep} included reaction \eqref{r1} (with an estimated activation energy of 30000 K) in her investigation of pulsation-driven shock chemistry in the inner ($<5R_*$) winds of IRC+10216, as well as the formation of several other diatomic hydrides by this mechanism. Though barriers on the order of $10^4$ K are prohibitively large in typical interstellar environments, the hot inner regions of CSEs represent one region where such processes may occur with anything other than negligible rates. They found that while this formation route of FeH is more efficient when considering shocks over a pure thermochemical equilibrium model, the derived abundances are still very low (less than the upper limits derived in this work) and fall off exponentially over the dust formation region of this CSE. Interestingly, experiments by \citet{chertihin_infrared_1995} demonstrated that the photoexcitation of Fe on surfaces could drive reaction \eqref{r1} at low temperatures, and an analogous process might occur on dust grains in the ISM, though not efficiently enough to produce observable quantities of FeH in this source. 

Our results likewise suggest that reactions \eqref{r2} - \eqref{r4} are similarly inefficient. For example, gas-phase radiative associations such as reaction \eqref{r2} can be astrochemically important \citep{herbst_radiative_1976,vasyunin_reactive_2013,balucani_formation_2015}, though as a general rule, emission of a photon is a much slower process than dissociation in the gas, particularly for small molecules like FeH -  thereby likely rendering this pathway negligible. Of the above reactions, \eqref{r4} stands out as perhaps the most promising mechanism in the outer envelope, since such ion-neutral reactions are often barrierless, and it is in these regions that photoionization of atomic Fe could be efficient enough to drive this process \citep{2009ApJ...697...68C}. However, previous studies suggest that FeH is not a major product of either, yielding instead $\mathrm{Fe^+} + \mathrm{H} + \mathrm{H_2}$ or $\mathrm{FeH^+} + \mathrm{H}$, for \eqref{r3} and \eqref{r4}, respectively \citep{harada_new_2010,Irikura:1990} - conclusions which our results here support.

\subsection{Si and P Hydrides in CSEs}
\label{sec:Si_P_chem}
Though larger hydrides are generally regarded as daughter species of smaller molecules like the ones targeted in this work, \cite{1995MNRAS.274..694M} proposed that in the hot cores of star forming regions, SiH could form through \ce{SiH4} desorbing off dust grains and undergoing hydrogen abstraction reactions. \ce{SiH4} is known to be abundant in both IRC+10216 and VY CMa \citep{2000ApJ...543..868M}, so if this process is occurring at a significant rate in either source, we would expect to see cold SiH emission coming from the intermediate CSEs. 

Such a formation mechanism for SiH would have to be quite efficient for us to have been able to detect this species in the targeted sources. This is because silicon monohydride reacts readily not only with atoms such as O, N, and C \citep{wakelam_2014_2015}, as well as \ce{SiH4} itself \citep{nemoto_electronic_1989} - but even with \ce{H2} \citep{walch_thermal_2001}. In their theoretical investigation of the \ce{SiH + H2} reaction, \citet{walch_thermal_2001} found that it occurs with a temperature-dependent 2$^\mathrm{nd}$ order rate coefficient given by $k(T) = 4.45\times10^{-11}e^{-3774/T}$ cm$^3$ s$^{-1}$. Using the temperature profiles derived by \citep{fonfria_detailed_2008}, we obtain rate coefficients of $k(900 K) = 7\times10^{-13}$ cm$^3$ s$^{-1}$ and $k(475 K) = 2\times10^{-14}$ cm$^3$ s$^{-1}$ for distances of 5 and 15 $R_*$, respectively. Though these rate coefficients are fairly low, given the large abundance of \ce{H2}, the overall reaction would be quite efficient. This fact is illustrated by expressing the 2$^\mathrm{nd}$ order rate coefficient in pseudo-1$^\mathrm{st}$ order form, assuming an \ce{H2} abundance of $n=10^7$ cm$^{-3}$, as $k^{'}(T) = 4.45\times10^{-4}e^{-3774/T}$ s$^{-1}$, which results in values of  $k^{'}(900 K) = 7\times10^{-6}$ s$^{-1}$ and $k^{'}(475 K) = 2\times10^{-7}$ s$^{-1}$, again, for distances of 5 and 15 $R_*$, respectively. Thus, the gas-phase abundance of SiH in this region of the CSE of both IRC+10216 and VY CMa should be quite low, unless there exists some more efficient method of forming it, which is not what our results suggest.

As for the production of phosphorous hydrides in CSEs, \cite{2014ApJ...790L..27A} found that \ce{PH3} accounts for about 2\% of the phosphorous content in IRC+10216, making it the second-most abundant gas-phase P-bearing molecule in the envelope (behind \ce{HCP}). Interestingly, they noted that this species might be formed efficiently on the dust grains that are so characteristic of CSEs. If that is the case then one of the logical precursors of \ce{PH3} would be PH. Of course, given the fairly large upper limits we have derived from our observational data, all we can say with confidence is that gas-phase PH does not appear to be a major reservoir of phosphorus in the sources targeted in this study.

Nevertheless, if H-addition on grains is indeed an efficient process in the inner regions of CSEs (ca. $5 < R_* < 20$), then our findings are still consistent with there being an possible underlying chemical connection between the non-detection of PH described here, and the previous detection of \ce{PH3} via the grain surface process

\begin{equation}
    \ce{PH + 2H -> PH3}. 
\end{equation}

\noindent
Though PH was not targeted in this work towards VY CMa, the same possible connection might likewise be relevant, though the higher O abundance than in IRC+10216 would make that source an even less hospitable environment for PH, given the efficiency of the reaction

\begin{equation}
    \ce{PH + O -> PO + H},
\end{equation}

\noindent
which has been estimated to proceed at approximately the collisional rate \citep{wakelam_2014_2015}.

Further comparisons with previous findings show that the upper limit abundances derived in this work are larger than the predicted abundances of \cite{2020A&A...637A..59A} by one and two orders of magnitude for \ce{SiH} and \ce{PH}, respectively. Therefore, we cannot confidently conclude that our non-detections are indicative of these species exhibiting non-equilibrium chemistry in IRC+10216. This observed agreement with inner wind models is interesting given that previous works have shown heavier Si- and P-bearing hydrides, namely silane (\ce{SiH4}) and phosphine (\ce{PH3}), are observed in stark \textit{overabundance} ($>6$ dex) relative to predictions of thermochemical equilibrium \citep{1993ApJ...406..199K,2014ApJ...790L..27A,2020A&A...637A..59A}. So while our results do not necessarily imply a preferred chemical pathway for the production of \ce{SiH} and \ce{PH} in CSEs, they are consistent with the notion that subsequent hydrogenation of such species occurs much more efficiently than a pure gas-phase equilibrium chemistry predicts.

\section{Conclusions}
We conducted a search for rotational transitions of SiH, PH, and FeH toward the CSEs of IRC+10216 and VY CMa using the GREAT spectrometer aboard the SOFIA observatory. We detected many high-$J$ transitions from a variety of molecules that have previously been studied in these sources; however, we saw no emission from any of our target species. We derive upper limit column densities and abundances for the targeted molecules in each source using the limiting sensitivities of our measurements along with known physical conditions of both envelopes. Though observational factors like beam dilution undoubtedly play a role in these non-detections, our upper limits reflect the inefficiency of predicted formation routes, and tendency for these hydrides to either be incorporated into dust grains or react with other gas-phase species abundant in these environments. Our findings also underscore the dichotomy between the observed abundances of diatomic hydrides relative to their heavier counterparts in CSEs, particularly in the cases of SiH and PH.
 
The presence and build-up mechanisms of hydrides in the inner regions of circumstellar material remain largely unconstrained by the present body of observational work. In order to better understand this process, there is a need for both high-sensitivity FIR measurements targeting new light hydrides like those in this work, in addition to high angular resolution sub-mm interferometric studies of known hydrides in CSEs like \ce{PH3}, \ce{H2O}, and \ce{NH3}.

\acknowledgements

The National Radio Astronomy Observatory is a facility of the National Science Foundation operated under cooperative agreement by Associated Universities, Inc. Support for B.A.M. during the initial portions of this work was provided by NASA through Hubble Fellowship grant \#HST-HF2-51396 awarded by the Space Telescope Science Institute, which is operated by the Association of Universities for Research in Astronomy, Inc., for NASA, under contract NAS5-26555. C. N. S. wishes to thank the Alexander von Humboldt Foundation for its support. A.M.B. acknowledges support from the Smithsonian Institution as a current Submillimeter Array (SMA) Fellow. 

Based on observations made with the NASA/DLR Stratospheric Observatory for Infrared Astronomy (SOFIA). SOFIA is jointly operated by the Universities Space Research Association, Inc. (USRA), under NASA contract NNA17BF53C, and the Deutsches SOFIA Institut (DSI) under DLR contract 50 OK 0901 to the University of Stuttgart. 

\bibliography{bibliography,chrisreferences}

\begin{thebibliography}{}
\expandafter\ifx\csname natexlab\endcsname\relax\def\natexlab#1{#1}\fi
\providecommand{\url}[1]{\href{#1}{#1}}
\providecommand{\dodoi}[1]{doi:~\href{http://doi.org/#1}{\nolinkurl{#1}}}
\providecommand{\doeprint}[1]{\href{http://ascl.net/#1}{\nolinkurl{http://ascl.net/#1}}}
\providecommand{\doarXiv}[1]{\href{https://arxiv.org/abs/#1}{\nolinkurl{https://arxiv.org/abs/#1}}}

\bibitem[{Adams(1941)}]{Adams:1941tj}
Adams, W.~S. 1941, The Astrophysical Journal, 93, 11

\bibitem[{{Ag{\'u}ndez} {et~al.}(2014){Ag{\'u}ndez}, {Cernicharo}, {Decin},
  {Encrenaz}, \& {Teyssier}}]{2014ApJ...790L..27A}
{Ag{\'u}ndez}, M., {Cernicharo}, J., {Decin}, L., {Encrenaz}, P., \&
  {Teyssier}, D. 2014, ApJ, 790, L27, \dodoi{10.1088/2041-8205/790/2/L27}

\bibitem[{{Ag{\'u}ndez} {et~al.}(2011){Ag{\'u}ndez}, {Cernicharo}, {Waters},
  {Decin}, {Encrenaz}, {Neufeld}, {Teyssier}, \&
  {Daniel}}]{2011A&A...533L...6A}
{Ag{\'u}ndez}, M., {Cernicharo}, J., {Waters}, L.~B.~F.~M., {et~al.} 2011,
  \aap, 533, L6, \dodoi{10.1051/0004-6361/201117578}

\bibitem[{{Ag{\'u}ndez} {et~al.}(2020){Ag{\'u}ndez}, {Mart{\'\i}nez}, {de
  Andres}, {Cernicharo}, \& {Mart{\'\i}n-Gago}}]{2020A&A...637A..59A}
{Ag{\'u}ndez}, M., {Mart{\'\i}nez}, J.~I., {de Andres}, P.~L., {Cernicharo},
  J., \& {Mart{\'\i}n-Gago}, J.~A. 2020, \aap, 637, A59,
  \dodoi{10.1051/0004-6361/202037496}

\bibitem[{{Alcolea} {et~al.}(2013){Alcolea}, {Bujarrabal}, {Planesas},
  {Teyssier}, {Cernicharo}, {De Beck}, {Decin}, {Dominik}, {Justtanont}, {de
  Koter}, {Marston}, {Melnick}, {Menten}, {Neufeld}, {Olofsson}, {Schmidt},
  {Sch{\"o}ier}, {Szczerba}, \& {Waters}}]{2013A&A...559A..93A}
{Alcolea}, J., {Bujarrabal}, V., {Planesas}, P., {et~al.} 2013, \aap, 559, A93,
  \dodoi{10.1051/0004-6361/201321683}

\bibitem[{{Amari} {et~al.}(1990){Amari}, {Anders}, {Virag}, \&
  {Zinner}}]{1990Natur.345..238A}
{Amari}, S., {Anders}, A., {Virag}, A., \& {Zinner}, E. 1990, \nat, 345, 238,
  \dodoi{10.1038/345238a0}

\bibitem[{{Avery} {et~al.}(1992){Avery}, {Amano}, {Bell}, {Feldman}, {Johns},
  {MacLeod}, {Matthews}, {Morton}, {Watson}, {Turner}, {Hayashi}, {Watt}, \&
  {Webster}}]{Avery:1992A}
{Avery}, L.~W., {Amano}, T., {Bell}, M.~B., {et~al.} 1992, \apjs, 83, 363,
  \dodoi{10.1086/191742}

\bibitem[{Balucani {et~al.}(2015)Balucani, Ceccarelli, \&
  Taquet}]{balucani_formation_2015}
Balucani, N., Ceccarelli, C., \& Taquet, V. 2015, Monthly Notices of the Royal
  Astronomical Society, 449, L16, \dodoi{10.1093/mnrasl/slv009}

\bibitem[{{Barklem} \& {Collet}(2016)}]{2016A&A...588A..96B}
{Barklem}, P.~S., \& {Collet}, R. 2016, \aap, 588, A96,
  \dodoi{10.1051/0004-6361/201526961}

\bibitem[{{Bernatowicz} {et~al.}(1991){Bernatowicz}, {Amari}, {Zinner}, \&
  {Lewis}}]{1991ApJ...373L..73B}
{Bernatowicz}, T.~J., {Amari}, S., {Zinner}, E.~K., \& {Lewis}, R.~S. 1991,
  \apjl, 373, L73, \dodoi{10.1086/186054}

\bibitem[{{Blake} {et~al.}(1985){Blake}, {Keene}, \&
  {Phillips}}]{1985ApJ...295..501B}
{Blake}, G.~A., {Keene}, J., \& {Phillips}, T.~G. 1985, \apj, 295, 501,
  \dodoi{10.1086/163394}

\bibitem[{Brown {et~al.}(2006)Brown, K{\"o}rsgen, Beaton, \&
  Evenson}]{Brown:2006gy}
Brown, J.~M., K{\"o}rsgen, H., Beaton, S.~P., \& Evenson, K.~M. 2006, The
  Journal of Chemical Physics, 124, 234309

\bibitem[{{Campins} \& {Ryan}(1989)}]{1989ApJ...341.1059C}
{Campins}, H., \& {Ryan}, E.~V. 1989, \apj, 341, 1059, \dodoi{10.1086/167563}

\bibitem[{Carroll \& McCormack(1972)}]{carroll_spectrum_1972}
Carroll, P.~K., \& McCormack, P. 1972, The Astrophysical Journal Letters, 177,
  L33, \dodoi{10.1086/181047}

\bibitem[{Carroll {et~al.}(1976)Carroll, McCormack, \&
  Oconnor}]{carroll_iron_1976}
Carroll, P.~K., McCormack, P., \& Oconnor, S. 1976, The Astrophysical Journal,
  208, 903, \dodoi{10.1086/154679}

\bibitem[{Caselli {et~al.}(1997)Caselli, Hartquist, \&
  Havnes}]{caselli_grain-grain_1997}
Caselli, P., Hartquist, T.~W., \& Havnes, O. 1997, Astronomy and Astrophysics,
  322, 296

\bibitem[{{Cernicharo} {et~al.}(2013){Cernicharo}, {Daniel}, {Castro-Carrizo},
  {Agundez}, {Marcelino}, {Joblin}, {Goicoechea}, \&
  {Gu{\'e}lin}}]{Cernicharo:2013A}
{Cernicharo}, J., {Daniel}, F., {Castro-Carrizo}, A., {et~al.} 2013, \apjl,
  778, L25, \dodoi{10.1088/2041-8205/778/2/L25}

\bibitem[{Cernicharo {et~al.}(2010)Cernicharo, Decin, Barlow, Ag{\'u}ndez,
  Royer, Vandenbussche, Wesson, Polehampton, De~Beck, Blommaert, Daniel,
  De~Meester, Exter, Feuchtgruber, Gear, Goicoechea, Gomez, Groenewegen,
  Hargrave, Huygen, Imhof, Ivison, Jean, Kerschbaum, Leeks, Lim, Matsuura,
  Olofsson, Posch, Regibo, Savini, Sibthorpe, Swinyard, Vandenbussche, \&
  Waelkens}]{Cernicharo:2010cv}
Cernicharo, J., Decin, L., Barlow, M.~J., {et~al.} 2010, Astronomy {\&}
  Astrophysics, 518, L136

\bibitem[{Cherchneff(2012)}]{Cherchneff:2012ep}
Cherchneff, I. 2012, Astronomy {\&} Astrophysics, 545, A12

\bibitem[{Chertihin \& Andrews(1995)}]{chertihin_infrared_1995}
Chertihin, G.~V., \& Andrews, L. 1995, The Journal of Physical Chemistry, 99,
  12131, \dodoi{10.1021/j100032a013}

\bibitem[{Codella {et~al.}(2010)Codella, Lefloch, Ceccarelli, Cernicharo, Caux,
  Lorenzani, Viti, Hily-Blant, Parise, Maret, Nisini, Caselli, Cabrit, Pagani,
  Benedettini, Boogert, Gueth, Melnick, Neufeld, Pacheco, Salez, Schuster,
  Bacmann, Baudry, Bell, Bergin, Blake, Bottinelli, Castets, Comito, Coutens,
  Crimier, Dominik, Demyk, Encrenaz, Falgarone, Fuente, Gerin, Goldsmith,
  Helmich, Hennebelle, Henning, Herbst, Jacq, Kahane, Kama, Klotz, Langer, Lis,
  Lord, Pearson, Phillips, Saraceno, Schilke, Tielens, van~der Tak, van~der
  Wiel, Vastel, Wakelam, Walters, Wyrowski, Yorke, Borys, Delorme, Kramer,
  Larsson, Mehdi, Ossenkopf, \& Stutzki}]{codella_CHESS_2010}
Codella, C., Lefloch, B., Ceccarelli, C., {et~al.} 2010, Astronomy and
  Astrophysics, 518, L112, \dodoi{10.1051/0004-6361/201014582}

\bibitem[{{Cordiner} \& {Millar}(2009)}]{2009ApJ...697...68C}
{Cordiner}, M.~A., \& {Millar}, T.~J. 2009, \apj, 697, 68,
  \dodoi{10.1088/0004-637X/697/1/68}

\bibitem[{Fonfría {et~al.}(2008)Fonfría, Cernicharo, Richter, \&
  Lacy}]{fonfria_detailed_2008}
Fonfría, J.~P., Cernicharo, J., Richter, M.~J., \& Lacy, J.~H. 2008, The
  Astrophysical Journal, 673, 445, \dodoi{10.1086/523882}

\bibitem[{Frisch {et~al.}(2009)Frisch, Trucks, Schlegel, Scuseria, Robb,
  Cheeseman, Scalmani, Barone, Mennucci, Petersson, Nakatsuji, Caricato, Li,
  Hratchian, Izmaylov, Bloino, Zheng, Sonnenberg, Hada, Ehara, Toyota, Fukuda,
  Hasegawa, Ishida, Nakajima, Honda, Kitao, Nakai, Vreven, Montgomery, Peralta,
  Ogliaro, Bearpark, Heyd, Brothers, Kudin, Staroverov, Kobayashi, Normand,
  Raghavachari, Rendell, Burant, Iyengar, Tomasi, Cossi, Rega, Millam, Klene,
  Knox, Cross, Bakken, Adamo, Jaramillo, Gomperts, Stratmann, Yazyev, Austin,
  Cammi, Pomelli, Ochterski, Martin, Morokuma, Zakrzewski, Voth, Salvador,
  Dannenberg, Dapprich, Daniels, {Farkas}, Foresman, Ortiz, Cioslowski, \&
  Fox}]{gaussian09B1}
Frisch, M.~J., Trucks, G.~W., Schlegel, H.~B., {et~al.} 2009, Gaussian 09,
  Revision B.01, Gaussian, Inc., Wallingford CT

\bibitem[{{Gehrz}(1989)}]{1989IAUS..135..445G}
{Gehrz}, R. 1989, in IAU Symposium, Vol. 135, Interstellar Dust, ed. L.~J.
  {Allamandola} \& A.~G.~G.~M. {Tielens}, 445

\bibitem[{Gong {et~al.}(2015)Gong, Henkel, Spezzano, Thorwirth, Menten,
  Wyrowski, Mao, \& Klein}]{Gong:2015ks}
Gong, Y., Henkel, C., Spezzano, S., {et~al.} 2015, Astronomy {\&} Astrophysics,
  574, A56

\bibitem[{Gordy \& Cook(1984)}]{Gordy:1984uy}
Gordy, W., \& Cook, R.~L. 1984, {Microwave Molecular Spectra}, 3rd edn. (New
  York: Wiley)

\bibitem[{Harada {et~al.}(2010)Harada, Herbst, \& Wakelam}]{harada_new_2010}
Harada, N., Herbst, E., \& Wakelam, V. 2010, The Astrophysical Journal, 721,
  1570, \dodoi{10.1088/0004-637X/721/2/1570}

\bibitem[{Herbst(1976)}]{herbst_radiative_1976}
Herbst, E. 1976, The Astrophysical Journal, 205, 94, \dodoi{10.1086/154253}

\bibitem[{Hollis {et~al.}(2004)Hollis, Jewell, Lovas, \&
  Remijan}]{Hollis:2004uh}
Hollis, J.~M., Jewell, P.~R., Lovas, F.~J., \& Remijan, A. 2004, The
  Astrophysical Journal, 613, L45

\bibitem[{Irikura {et~al.}(1990)Irikura, Goddard, \& Beuchamp}]{Irikura:1990}
Irikura, K., Goddard, W., \& Beuchamp, J. 1990, International Journal of Mass
  Spectrometry and Ion Processes, 99, 213

\bibitem[{{Jones} {et~al.}(1996){Jones}, {Tielens}, \&
  {Hollenbach}}]{tielens_grain_1996}
{Jones}, A.~P., {Tielens}, A.~G.~G.~M., \& {Hollenbach}, D.~J. 1996, \apj, 469,
  740, \dodoi{10.1086/177823}

\bibitem[{Jones {et~al.}(1994)Jones, Tielens, Hollenbach, \&
  McKee}]{jones_grain_1994}
Jones, A.~P., Tielens, A. G. G.~M., Hollenbach, D.~J., \& McKee, C.~F. 1994,
  The Astrophysical Journal, 433, 797, \dodoi{10.1086/174689}

\bibitem[{{Kami{\'n}ski} {et~al.}(2013{\natexlab{a}}){Kami{\'n}ski},
  {Gottlieb}, {Young}, {Menten}, \& {Patel}}]{2013ApJS..209...38K}
{Kami{\'n}ski}, T., {Gottlieb}, C.~A., {Young}, K.~H., {Menten}, K.~M., \&
  {Patel}, N.~A. 2013{\natexlab{a}}, \apjs, 209, 38,
  \dodoi{10.1088/0067-0049/209/2/38}

\bibitem[{{Kami{\'n}ski} {et~al.}(2013{\natexlab{b}}){Kami{\'n}ski}, {Schmidt},
  \& {Menten}}]{2013A&A...549A...6K}
{Kami{\'n}ski}, T., {Schmidt}, M.~R., \& {Menten}, K.~M. 2013{\natexlab{b}},
  \aap, 549, A6, \dodoi{10.1051/0004-6361/201220650}

\bibitem[{Kami{\'{n}}ski {et~al.}(2013)Kami{\'{n}}ski, Gottlieb, Menten, Patel,
  Young, Br{\"u}nken, M{\"u}ller, McCarthy, Winters, \&
  Decin}]{Kaminski:2013gk}
Kami{\'{n}}ski, T., Gottlieb, C.~A., Menten, K.~M., {et~al.} 2013, Astronomy
  {\&} Astrophysics, 551, 113

\bibitem[{{Keady} \& {Ridgway}(1993)}]{1993ApJ...406..199K}
{Keady}, J.~J., \& {Ridgway}, S.~T. 1993, \apj, 406, 199,
  \dodoi{10.1086/172431}

\bibitem[{{Mackay}(1995)}]{1995MNRAS.274..694M}
{Mackay}, D.~D.~S. 1995, MNRAS, 274, 694, \dodoi{10.1093/mnras/274.3.694}

\bibitem[{{Massalkhi} {et~al.}(2019){Massalkhi}, {Ag{\'u}ndez}, \&
  {Cernicharo}}]{2019A&A...628A..62M}
{Massalkhi}, S., {Ag{\'u}ndez}, M., \& {Cernicharo}, J. 2019, \aap, 628, A62,
  \dodoi{10.1051/0004-6361/201935069}

\bibitem[{{Massalkhi} {et~al.}(2018){Massalkhi}, {Ag{\'u}ndez}, {Cernicharo},
  {Velilla Prieto}, {Goicoechea}, {Quintana-Lacaci}, {Fonfr{\'\i}a}, {Alcolea},
  \& {Bujarrabal}}]{Massalkhi:611A29M}
{Massalkhi}, S., {Ag{\'u}ndez}, M., {Cernicharo}, J., {et~al.} 2018, \aap, 611,
  A29, \dodoi{10.1051/0004-6361/201732038}

\bibitem[{{Matthews} {et~al.}(2015){Matthews}, {G{\'e}rard}, \& {Le
  Bertre}}]{2015MNRAS.449..220M}
{Matthews}, L.~D., {G{\'e}rard}, E., \& {Le Bertre}, T. 2015, \mnras, 449, 220,
  \dodoi{10.1093/mnras/stv263}

\bibitem[{Mauron \& Huggins(2010)}]{Mauron:2010dk}
Mauron, N., \& Huggins, P.~J. 2010, A{\&}A, 513, A31

\bibitem[{{McGuire}(2018)}]{2018ApJS..239...17M}
{McGuire}, B.~A. 2018, \apjs, 239, 17, \dodoi{10.3847/1538-4365/aae5d2}

\bibitem[{{Meyer} \& {Roth}(1991)}]{1991ApJ...376L..49M}
{Meyer}, D.~M., \& {Roth}, K.~C. 1991, \apjl, 376, L49, \dodoi{10.1086/186100}

\bibitem[{{Monnier} {et~al.}(2000){Monnier}, {Danchi}, {Hale}, {Tuthill}, \&
  {Townes}}]{2000ApJ...543..868M}
{Monnier}, J.~D., {Danchi}, W.~C., {Hale}, D.~S., {Tuthill}, P.~G., \&
  {Townes}, C.~H. 2000, \apj, 543, 868, \dodoi{10.1086/317127}

\bibitem[{Mould \& Wyckoff(1978{\natexlab{a}})}]{Mould:1978}
Mould, J., \& Wyckoff, S. 1978{\natexlab{a}}, Monthly Notices of the Royal
  Astronomical Society, 182, 63

\bibitem[{Mould \& Wyckoff(1978{\natexlab{b}})}]{mould_iron_1978}
Mould, J.~R., \& Wyckoff, S. 1978{\natexlab{b}}, Monthly Notices of the Royal
  Astronomical Society, 182, 63, \dodoi{10.1093/mnras/182.1.63}

\bibitem[{{M{\"u}ller} {et~al.}(2005){M{\"u}ller}, {Schl{\"o}der}, {Stutzki},
  \& {Winnewisser}}]{2005JMoSt.742..215M}
{M{\"u}ller}, H. S.~P., {Schl{\"o}der}, F., {Stutzki}, J., \& {Winnewisser}, G.
  2005, Journal of Molecular Structure, 742, 215,
  \dodoi{10.1016/j.molstruc.2005.01.027}

\bibitem[{Nemoto {et~al.}(1989)Nemoto, Suzuki, Nakamura, Shibuya, \&
  Obi}]{nemoto_electronic_1989}
Nemoto, M., Suzuki, A., Nakamura, H., Shibuya, K., \& Obi, K. 1989, Chemical
  Physics Letters, 162, 467, \dodoi{10.1016/0009-2614(89)87009-5}

\bibitem[{{Neufeld} {et~al.}(1997){Neufeld}, {Zmuidzinas}, {Schilke}, \&
  {Phillips}}]{1997ApJ...488L.141N}
{Neufeld}, D.~A., {Zmuidzinas}, J., {Schilke}, P., \& {Phillips}, T.~G. 1997,
  \apjl, 488, L141, \dodoi{10.1086/310942}

\bibitem[{{Neufeld} {et~al.}(2012){Neufeld}, {Falgarone}, {Gerin}, {Godard},
  {Herbst}, {Pineau des For{\^e}ts}, {Vasyunin}, {G{\"u}sten}, {Wiesemeyer}, \&
  {Ricken}}]{2012A&A...542L...6N}
{Neufeld}, D.~A., {Falgarone}, E., {Gerin}, M., {et~al.} 2012, \aap, 542, L6,
  \dodoi{10.1051/0004-6361/201218870}

\bibitem[{Nordh {et~al.}(1977)Nordh, Lindgren, \& Wing}]{nordh_proposed_1977}
Nordh, H.~L., Lindgren, B., \& Wing, R.~F. 1977, Astronomy and Astrophysics,
  56, 1

\bibitem[{{Patel} {et~al.}(2011){Patel}, {Young}, {Gottlieb}, {Thaddeus},
  {Wilson}, {Menten}, {Reid}, {McCarthy}, {Cernicharo}, {He}, {Br{\"u}nken},
  {Trung}, \& {Keto}}]{2011ApJS..193...17P}
{Patel}, N.~A., {Young}, K.~H., {Gottlieb}, C.~A., {et~al.} 2011, \apjs, 193,
  17, \dodoi{10.1088/0067-0049/193/1/17}

\bibitem[{{Rogantini} {et~al.}(2019){Rogantini}, {Costantini}, {Zeegers}, {de
  Vries}, {Mehdipour}, {de Groot}, {Mutschke}, {Psaradaki}, \&
  {Waters}}]{2019A&A...630A.143R}
{Rogantini}, D., {Costantini}, E., {Zeegers}, S.~T., {et~al.} 2019, \aap, 630,
  A143, \dodoi{10.1051/0004-6361/201935883}

\bibitem[{{Sahai} \& {Wannier}(1992)}]{1992ApJ...394..320S}
{Sahai}, R., \& {Wannier}, P.~G. 1992, \apj, 394, 320, \dodoi{10.1086/171585}

\bibitem[{Schilke {et~al.}(1997)Schilke, Walmsley, Pineau~des Forets, \&
  Flower}]{schilke_SiO_1997}
Schilke, P., Walmsley, C.~M., Pineau~des Forets, G., \& Flower, D.~R. 1997,
  Astronomy and Astrophysics, 321, 293

\bibitem[{{Shenoy} {et~al.}(2016){Shenoy}, {Humphreys}, {Jones}, {Marengo},
  {Gehrz}, {Helton}, {Hoffmann}, {Skemer}, \& {Hinz}}]{Shenoy:2016AJ}
{Shenoy}, D., {Humphreys}, R.~M., {Jones}, T.~J., {et~al.} 2016, \aj, 151, 51,
  \dodoi{10.3847/0004-6256/151/3/51}

\bibitem[{{Smith} {et~al.}(2001){Smith}, {Humphreys}, {Davidson}, {Gehrz},
  {Schuster}, \& {Krautter}}]{Smith:2001A}
{Smith}, N., {Humphreys}, R.~M., {Davidson}, K., {et~al.} 2001, \aj, 121, 1111,
  \dodoi{10.1086/318748}

\bibitem[{Sofia {et~al.}(1994)Sofia, Cardelli, \& Savage}]{sofia_abundant_1994}
Sofia, U.~J., Cardelli, J.~A., \& Savage, B.~D. 1994, The Astrophysical
  Journal, 430, 650, \dodoi{10.1086/174438}

\bibitem[{{Swings} \& {Rosenfeld}(1937)}]{1937ApJ....86..483S}
{Swings}, P., \& {Rosenfeld}, L. 1937, \apj, 86, 483, \dodoi{10.1086/143880}

\bibitem[{{Tenenbaum} {et~al.}(2010){Tenenbaum}, {Dodd}, {Milam}, {Woolf}, \&
  {Ziurys}}]{2010ApJS..190..348T}
{Tenenbaum}, E.~D., {Dodd}, J.~L., {Milam}, S.~N., {Woolf}, N.~J., \& {Ziurys},
  L.~M. 2010, \apjs, 190, 348, \dodoi{10.1088/0067-0049/190/2/348}

\bibitem[{Tielens {et~al.}(1994)Tielens, McKee, Seab, \&
  Hollenbach}]{tielens_physics_1994}
Tielens, A. G. G.~M., McKee, C.~F., Seab, C.~G., \& Hollenbach, D.~J. 1994, The
  Astrophysical Journal, 431, 321, \dodoi{10.1086/174488}

\bibitem[{Vasyunin \& Herbst(2013)}]{vasyunin_reactive_2013}
Vasyunin, A.~I., \& Herbst, E. 2013, The Astrophysical Journal, 769, 34,
  \dodoi{10.1088/0004-637X/769/1/34}

\bibitem[{Wakelam {et~al.}(2015)Wakelam, Loison, Herbst, Pavone, Bergeat,
  Béroff, Chabot, Faure, Galli, Geppert, Gerlich, Gratier, Harada, Hickson,
  Honvault, Klippenstein, Picard, Nyman, Ruaud, Schlemmer, Sims, Talbi,
  Tennyson, \& Wester}]{wakelam_2014_2015}
Wakelam, V., Loison, J.-C., Herbst, E., {et~al.} 2015, The Astrophysical
  Journal Supplement Series, 217, 20

\bibitem[{Walch \& Dateo(2001)}]{walch_thermal_2001}
Walch, S.~P., \& Dateo, C.~E. 2001, The Journal of Physical Chemistry A, 105,
  2015, \dodoi{10.1021/jp003559u}

\bibitem[{Walmsley {et~al.}(2002)Walmsley, Bachiller, Pineau~des Forêts, \&
  Schilke}]{walmsley_detection_2002}
Walmsley, C.~M., Bachiller, R., Pineau~des Forêts, G., \& Schilke, P. 2002,
  The Astrophysical Journal Letters, 566, L109, \dodoi{10.1086/339694}

\bibitem[{{Waters} {et~al.}(1998){Waters}, {Beintema}, {Zijlstra}, {de Koter},
  {Molster}, {Bouwman}, {de Jong}, {Pottasch}, \& {de
  Graauw}}]{1998A&A...331L..61W}
{Waters}, L.~B.~F.~M., {Beintema}, D.~A., {Zijlstra}, A.~A., {et~al.} 1998,
  \aap, 331, L61.
\newblock \doarXiv{astro-ph/9802289}

\bibitem[{Wing {et~al.}(1977)Wing, Cohen, \& Brault}]{wing_confirmation_1977}
Wing, R.~F., Cohen, J., \& Brault, J.~W. 1977, The Astrophysical Journal, 216,
  659, \dodoi{10.1086/155507}

\bibitem[{Wing \& Ford(1969)}]{wing_infrared_1969}
Wing, R.~F., \& Ford, Jr., W.~K. 1969, Publications of the Astronomical Society
  of the Pacific, 81, 527, \dodoi{10.1086/128814}

\bibitem[{Zack {et~al.}(2011)Zack, Halfen, \& Ziurys}]{Zack:2011jx}
Zack, L.~N., Halfen, D.~T., \& Ziurys, L.~M. 2011, The Astrophysical Journal,
  733, L36

\bibitem[{{Zhang} {et~al.}(2017){Zhang}, {Zhu}, {Li}, {Chen}, {Wang}, \&
  {Zhang}}]{Zhang:2017A}
{Zhang}, X.-Y., {Zhu}, Q.-F., {Li}, J., {et~al.} 2017, \aap, 606, A74,
  \dodoi{10.1051/0004-6361/201730791}

\bibitem[{Ziurys(2006)}]{Ziurys:2006n}
Ziurys, L. 2006, PNAS, 103, 12274

\bibitem[{{Ziurys} {et~al.}(2007{\natexlab{a}}){Ziurys}, {Milam}, {Apponi}, \&
  {Woolf}}]{2007Natur.447.1094Z}
{Ziurys}, L.~M., {Milam}, S.~N., {Apponi}, A.~J., \& {Woolf}, N.~J.
  2007{\natexlab{a}}, \nat, 447, 1094, \dodoi{10.1038/nature05905}

\bibitem[{{Ziurys} {et~al.}(2007{\natexlab{b}}){Ziurys}, {Milam}, {Apponi}, \&
  {Woolf}}]{Ziurys:2007A}
---. 2007{\natexlab{b}}, \nat, 447, 1094, \dodoi{10.1038/nature05905}

\bibitem[{{Ziurys} {et~al.}(2018){Ziurys}, {Schmidt}, \&
  {Bernal}}]{2018ApJ...856..169Z}
{Ziurys}, L.~M., {Schmidt}, D.~R., \& {Bernal}, J.~J. 2018, \apj, 856, 169,
  \dodoi{10.3847/1538-4357/aaafc6}

\end{thebibliography}
\bibliographystyle{aasjournal}

\end{document}